\documentclass[aps,pra,preprintnumbers,showpacs,tightenlines]{revtex4}
\usepackage{amssymb}
\usepackage{amsmath}
\usepackage{graphicx}
\usepackage{epsfig}
\usepackage{subfigure}
\usepackage{amsfonts}
\usepackage{CJK}

\begin{document}

\title{Multiplex-controlled phase gate with qubits distributed in a multi-cavity system}

\author{Biaoliang Ye$^{1}$, Zhen-Fei Zheng$^{2}$, and Chui-Ping Yang$^{1,3}$}
\email{yangcp@hznu.edu.cn}
\address{$^1$Quantum Information Research Center, Shangrao Normal University, Shangrao 334001, China}
\address{$^2$Key Laboratory of Quantum Information, University of Science and Technology of China, Heifei 230026, China}
\address{$^3$Department of Physics, Hangzhou Normal University, Hangzhou 310036, China}

\date{\today}

\begin{abstract}
We present a way to realize a multiplex-controlled phase gate of $n-1$
control qubits simultaneously controlling one target qubit, with $n$ qubits
distributed in $n$ different cavities. This multiqubit gate is implemented by using $n$ qutrits (
three-level natural or artificial atoms) placed in $n$ different cavities, which are coupled to an auxiliary qutrit.
Here, the two logic states of a qubit are represented by the two lowest levels of a qutrit
placed in a cavity. We show that this $n$-qubit controlled phase gate can be realized using only $2n+2$ basic operations,
i.e., the number of required basic operations only increases
\textit{linearly} with the number $n$ of qubits. Since each basic operation
employs the qutrit-cavity or qutrit-pulse resonant interaction, the gate can
be fast implemented when the number of qubits is not large. Numerical
simulations show that a three-qubit controlled phase gate, which is executed on three
qubits distributed in three different cavities, can be high-fidelity implemented
by using a circuit QED system. This proposal is quite general and can be applied
to a wide range of physical systems, with atoms, NV centers, quantum dots,
or various superconducting qutrits distributed in different cavities. Finally,
this method can be applied to implement a multiqubit controlled phase gate with atoms using a cavity.
A detailed discussion on implementing a three-qubit controlled phase gate with atoms and one cavity is presented.
\end{abstract}
\pacs{03.67.Bg, 42.50.Dv, 85.25.Cp}\maketitle
\date{\today }

\begin{center}
\textbf{I. INTRODUCTION AND MOTIVATION}
\end{center}

Cavity or circuit QED has drawn much attention in the growing feild of
quantum information processing (QIP).  Large-scale QIP will most likely need
a large number of qubits, and placing all of them in a single cavity may
cause practical problems such as decreasing the qubit-cavity coupling
strength and increasing the cavity decay rate. Moreover, when compared to
QIP with qubits in a single cavity (Fig.~1), the size of QIP with qubits in multiple
cavities can be much larger, since the number $n\times m$ of qubits placed
in $n$ cavities is $n$ times the number $m$ of qubits placed in a single
cavity provided the number of qubits in each cavity is $m$. Large-scale QIP
may require quantum networks consisting of many cavities, each hosting and
coupled to multiple qubits. In such an architecture, manipulation of quantum
states will occur not only among qubits in the same cavity, \textit{but also
among qubits distributed in different cavities}. Thus, how to implement
quantum gates on qubits distributed in different cavities becomes necessary
and important.

Multiqubit gates are a crucial element in QIP. Generally speaking, there are
two types of important multiqubit gates, which have drawn much attention
during the past years. The first type of multiqubit gate consists of
multiple control qubits simultaneously controlling on a single target qubit
[1,2], while the second type of multiqubit gate contains a single qubit
simultaneously controlling multiple target qubits [3,4]. It is well known
that these two types of multiqubit gates are of significance in QIP. For
instances, they have applications in quantum algorithms [5--7], quantum
Fourier transform [1], error correction [8--10], quantum cloning [11], and
entanglement preparation [12].

\begin{figure}[tbp]
\begin{center}
\includegraphics[bb=186 442 378 612, width=7.0 cm, clip]{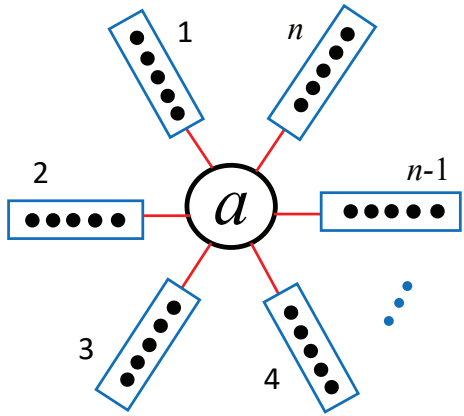} \vspace*{%
-0.08in}
\end{center}
\caption{(Color online) Setup of $n$ cavities connected by a coupler $a$ (a circle at the
center). A square represents a cavity, while a dark dot labels a qubit located in each cavity.
For clarity, only five qubits in each cavity are shown. In such an architecture, manipulation of quantum
states will occur not only among qubits in the same cavity, but also among qubits distributed in different cavities.
An important quantum gate among qubits in different cavities is a multiqubit controlled gate
with qubits distributed in different cavities. The qubits, which are not involved in the gate operation,
can be made to be decoupled to their respective cavities by adjusting the level spacings of qubits or loading the qubits out of the cavities.}
\label{fig:1}
\end{figure}

Motivated by the above, the focus of this work is on the direct implementation of the first type of
multiqubit gate (i.e., a multi-control-qubit gate with multiple control
qubits simultaneously controlling a target qubit) with qubits distributed in
a multi-cavity system. It is known that this multiqubit controlled gate can
in principle be constructed by using single-qubit and two-qubit basic gates.
However, when using the conventional gate-decomposing protocols to construct
this multiqubit controlled gate [2,13,14], the number of required
single-qubit and two-qubit gates increases drastically with the number of
qubits. Thus, the gate implementation becomes very complex, even in the case
of assuming a universal single-qubit or two-qubit gate could be realized
with a single basic operation. As a result, the gate operation time would be
quite long and thus the gate fidelity would be significantly deteriorated by
decoherence. Hence, it is worthwhile to seek efficient approaches to realize
this type of multi-control-qubit gates.

To begin with, let us give a brief review on the physical realization of multiqubit quantum gates.
For the past years, several schemes for
realizing three-qubit Toffoli gates have been proposed with neutral atoms in
an optical lattice [15] or hybrid atom-photon qubits [16]. In addition,
experimental realization of a controlled phase gate in a three qubit NMR
quantum system [17] and a three-qubit Toffoli gate with superconducting
qubits [18] has been reported. On the other hand, based on cavity or circuit
QED, many theoretical proposals have been presented for directly realizing
not only multi-control-qubit gates [19-29] but also multi-target-qubit gates
[4,30-33], in various physical qubits. Refs. [19,21-24] discussed how to
implement a multi-control-qubit gate with natural or artificial atoms
coupled to a single cavity. Refs. [20,25] considered how to realize a
multi-control-qubit gate with trapped ions. Refs. [26,27] addressed how to
implement a three-qubit control gate with three quantum dots embedded in
three spatially-separated cavities with assistance of linear optical devices
and photon detections. Refs. [28,29] proposed how to accomplish a
multi-control-qubit gate with flying photonic qubits distributed in
different cavities coupled to a single qutrit. Refs. [4,30-33] discussed on
how to implement a multi-target-qubit gate with natural or artificial atoms
placed in or coupled to a single cavity.

Different from previous works [4,19-33], in the following we will propose a method to
implement an $n$-qubit controlled phase gate of $n-1$ controlled qubits
simultaneously controlling a single target qubit, with $n$ qubits
in $n$ different cavities coupled to an auxiliary qutrit [Fig. 2(a)].
The two logic states of each qubit here are represented by the two lowest levels of a qutrit (a natural or
artificial three-level atom) placed in a cavity [Fig.~2(a)]. As shown below, this proposal
has the following advantages: (i) Only one auxiliary qutrit is needed and no
other auxiliary system is required; (ii) Since no measurement is needed, the
gate realization is deterministic; (iii) Only resonant interaction is used,
thus the gate operation can be performed fast for a small number of qubits;
and (iv) Because of only $2n+2$ basic operations are needed, the number of
basic operations only increases \textit{linearly} with the number $n$ of
qubits; thus the gate procedure is significantly simplified when compared
with the conventional gate-decomposing protocols. To the best of our
knowledge, this proposal was not reported before.

Finally, it is noted that this method can be applied to implement a multiqubit controlled
phase gate with atoms using a cavity. As an example, we will explicitly show
how to use the present method to implement a three-qubit controlled phase
gate with atoms by loading atoms into or moving atoms out of a cavity.

We stress that the main purpose of this work is on the implementation of a multiqubit controlled
gate with qubits distributed in a multi-cavity system instead of a single cavity. As discussed at the beginning,
implementing quantum gates on qubits distributed in different cavities is important in
large-scale QIP.

This paper is organized as follows. In Sec. II, we review the basic theory.
In Sec. III, we explicitly show how to realize a three-qubit controlled
phase gate with three qutrits distributed in three different cavities. In
Sec. IV, we then discuss how the method can be generalized to implement a $n$%
-qubit controlled phase gate with $n$ qutrits distributed in $n$ different
cavities. In Sec. V, based on a circuit QED system, we discuss the
experimental feasibility of implementing the proposed gate for a three-qubit
case. In Sec. VI, we show how to apply the method to realize a three-qubit
controlled phase gate with atoms using one cavity. A concluding summary is
given in Sec. VII.

\begin{figure}[tbp]
\begin{center}
\includegraphics[bb=35 401 566 629, width=12.5 cm, clip]{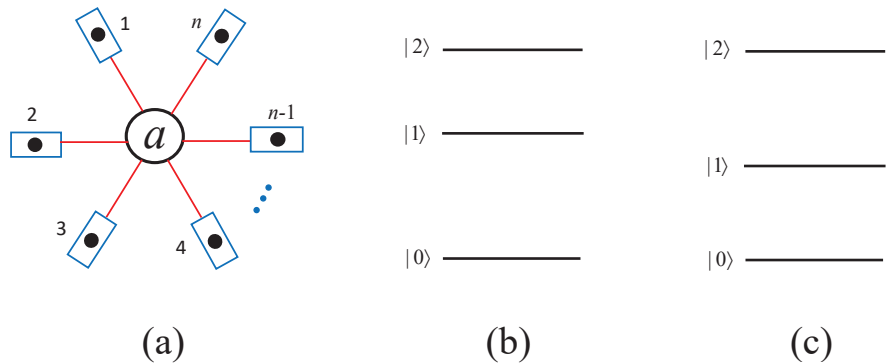} \vspace*{%
-0.08in}
\end{center}
\caption{(Color online) (a) Diagram of a coupler qutrit $a$ (a circle at the
center) and $n$ cavities ($1,2,...,n$) each hosting a qutrit. A square
represents a cavity, while a dark dot labels a qutrit placed in each cavity.
(b) A qutrit with three levels $\left\vert 0\right\rangle ,$ $\left\vert
1\right\rangle $ and $\left\vert 2\right\rangle $, for which the level
spacing between the upper two levels is smaller than that between the two
lowest levels. (c) A qutrit, whose level spacing between the upper two
levels is greater than that between the two lowest levels. As shown in next
section, the $\left\vert 0\right\rangle \leftrightarrow $ $\left\vert
1\right\rangle $ and $\left\vert 1\right\rangle \leftrightarrow \left\vert
2\right\rangle $ transitions are required by the gate implementation while
the $\left\vert 0\right\rangle \leftrightarrow $ $\left\vert 2\right\rangle $
transition is not necessary. Note that the level structure in (b) applies to
natural atoms, quantum dots, superconducting phase, transmon, and Xmon
qutrits; while the level structure in (c) is available in superconducting
charge qutrits, flux qutrits, nitrogen-vacancy centers, etc.}
\label{fig:2}
\end{figure}

\begin{center}
\textbf{II. BASIC THEORY}
\end{center}

Our multi-qubit gate is realized with a setup illustrated in Fig. 2(a),
where all cavities are coupled to an auxiliary qutrit and each cavity hosts
a qutrit. The three levels of each qutrit are denoted as $\left\vert
0\right\rangle ,$ $\left\vert 1\right\rangle $ and $\left\vert
2\right\rangle $, respectively [Fig.~2(b) and Fig.~2(c)]. As an example, our
following presentation starts with qutrits having the level structure
depicted in Fig.~2(b). However, we stress that the method presented below
for the gate implementation can directly apply to the qutrits with the level
structures shown in Fig.~2(c), because for this type of level structure the
required Hamiltonians given below can also be obtained.

The qutrit located in cavity $l$ is labelled as qutrit $l$ ($l=1,2,...,n$),
while the auxiliary qutrit, which is coupled to the $n$ cavities, is denoted
as qutrit $a$. As shown in next section, the gate implementation requires:
(i) A classical pulse resonantly interacting with the $\left\vert
1\right\rangle \leftrightarrow $ $\left\vert 2\right\rangle $ transition for
each of qutrits ($2,3,...,n$); (ii) Each cavity resonantly interacting with
the $\left\vert 0\right\rangle \leftrightarrow $ $\left\vert 1\right\rangle $
transition of qutrit $a$; and (iii) Each cavity simultaneously and
resonantly interacting with the $\left\vert 0\right\rangle \leftrightarrow $
$\left\vert 1\right\rangle $ transition of two qutrits (i.e., qutrit $a$ and
qutrit $l$ in cavity $l$). In the following, we will give a brief
introduction to the state evolution under these types of interaction.

\begin{figure}[tbp]
\begin{center}
\includegraphics[bb=92 174 497 352, width=15.0 cm, clip]{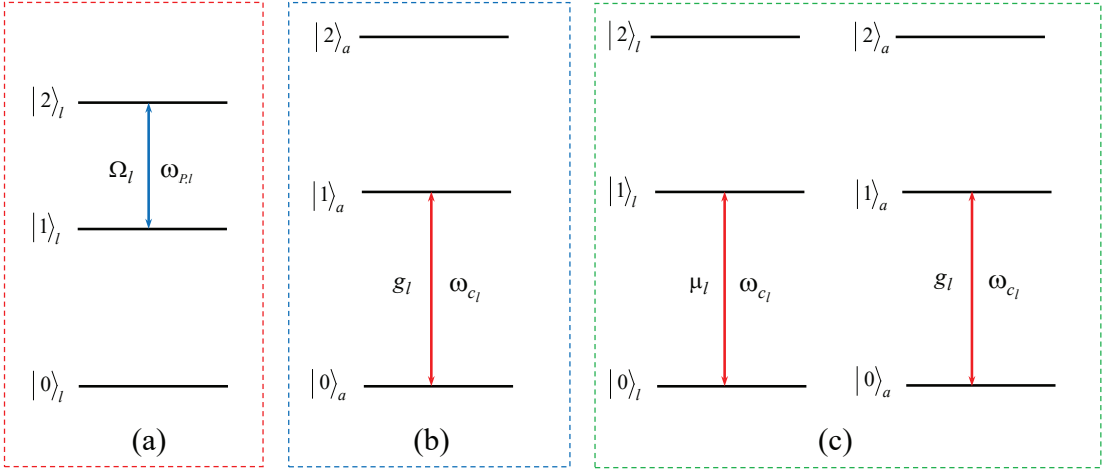} \vspace*{%
-0.08in}
\end{center}
\caption{(Color online) (a) Illustration of a classical pulse resonant with
the $\left\vert 1\right\rangle \leftrightarrow \left\vert 2\right\rangle $
transition of qutrit $l$. Here, $\Omega_{l}$ is the pulse Rabi frequency.
(b) Illustration of cavity $l$ resonant with the $\left\vert 0\right\rangle
\leftrightarrow \left\vert 1\right\rangle $ transition of qutrit $a$, with a
coupling constant $g_l$. (c) Illustration of cavity $l$ simultaneously
resonant with the $\left\vert 0\right\rangle \leftrightarrow \left\vert
1\right\rangle $ transition of qutrits $l$ and $a$, with a coupling constant
$\protect\mu_l$ for qutrit $l$ and a coupling constant $g_l$ for qutrit $a$.
Note that the level structure of qutrit $l$ in (a) is different from that in
(b) and (c), because qutrit $l$ in (a) is decoupled from cavity $l$ during
the pulse while the level spacings of qutrits in (b) and (c) are adjusted
such that the $\left\vert 0\right\rangle \leftrightarrow \left\vert
1\right\rangle $ transition is resonant with cavity $l$. Each horizontal
line represents the energy level of qutrit $l$ or qutrit $a$. A blue
double-arrow vertical line in (a) represents the frequency $\protect\omega%
_{p,l}$ of the pulse applied to qutrit $l$, while a red double-arrow
vertical line in (b) and (c) represents the frequency $\protect\omega%
_{c_{l}} $ of cavity $l$.}
\label{fig:3}
\end{figure}

\begin{center}
\textbf{A. Qutrit driven by a classical pulse}
\end{center}

Let us consider that a classical pulse is applied to qutrit $l$ ($%
l=2,3,...,n $). It can be shown that if the pulse is resonant with the $%
\left\vert 1\right\rangle \leftrightarrow \left\vert 2\right\rangle $
transition but far-off resonant with (decoupled from) the $\left\vert
0\right\rangle \leftrightarrow \left\vert 1\right\rangle $ transition and
the $\left\vert 0\right\rangle \leftrightarrow \left\vert 2\right\rangle $
transition of the qutrit [Fig.~3(a)], then the interaction Hamiltonian in
the interaction picture and after making a rotating-wave approximation
(RWA)\ is given by
\begin{equation}
H_{\mathrm{I}_{1}}=\hbar \left( \Omega _{l}e^{i\phi _{l}}\left\vert
1\right\rangle _{l}\left\langle 2\right\vert +\text{h.c.}\right) ,
\end{equation}%
where $\phi _{l}$ and $\Omega _{l}$ are the initial phase and the Rabi
frequency of the pulse, respectively. From the Hamiltonian (1), it is
straightforward to show that a pulse of duration $t$ results in the
following rotation
\begin{eqnarray}
\left\vert 1\right\rangle _{l} &\rightarrow &\cos \Omega _{l}t\left\vert
1\right\rangle _{l}-ie^{-i\phi _{l}}\sin \Omega _{l}t\left\vert
2\right\rangle _{l},  \notag \\
\left\vert 2\right\rangle _{l} &\rightarrow &-ie^{i\phi _{l}}\sin \Omega
_{l}t\left\vert 1\right\rangle _{l}+\cos \Omega _{l}t\left\vert
2\right\rangle _{l}.
\end{eqnarray}

\begin{center}
\textbf{B. Qutrit coupled to a single cavity}
\end{center}

Consider qutrit $a$ coupled to cavity $l$ ($l=1,2,...,n$). Suppose that
cavity $l$ is resonantly coupled to the $\left\vert 0\right\rangle
\leftrightarrow \left\vert 1\right\rangle $ transition while highly detunned
(decoupled) from the transitions between other levels of the qutrit
[Fig.~3(b)]. Under this consideration, the interaction Hamiltonian in the
interaction picture and after the RWA, can be written as
\begin{equation}
H_{\mathrm{I}_{2}}=\hbar \left( g_{_{l}}a_{l}^{+}\left\vert 0\right\rangle
_{a}\left\langle 1\right\vert +\text{h.c.}\right) ,
\end{equation}%
where $a_{l}^{+}$ ($a_{l}$) is the creation (annihilation) operator of
cavity $l$ while $g_{l}$ is the coupling constant between cavity $l$ and the
$\left\vert 0\right\rangle \leftrightarrow \left\vert 1\right\rangle $
transition of qutrit $a$.

Under the Hamiltonian (3), one can obtain the following state evolution:
\begin{eqnarray}
\left\vert 0\right\rangle _{a}\left\vert 0\right\rangle _{c_{l}}
&\rightarrow &\left\vert 0\right\rangle _{a}\left\vert 0\right\rangle
_{c_{l}},  \notag \\
\left\vert 1\right\rangle _{a}\left\vert 0\right\rangle _{c_{l}}
&\rightarrow &-i\sin g_{l}t\left\vert 0\right\rangle _{a}\left\vert
1\right\rangle _{c_{l}}+\cos g_{l}t\left\vert 1\right\rangle _{a}\left\vert
0\right\rangle _{c_{l}},
\end{eqnarray}%
where $\left\vert 0\right\rangle _{c_{l}}$ is the vacuum state of cavity $l$
while $\left\vert 1\right\rangle _{c_{l}}$ is the single-photon state of
cavity $l$. \textbf{\ }

\begin{center}
\textbf{C. Two qutrits coupled to a single cavity}
\end{center}

Consider two qutrits $l$ and $a$ coupled to cavity $l$ ($l=1,2,...,n$).
Suppose that the cavity is resonantly coupled to the $\left\vert
0\right\rangle \leftrightarrow \left\vert 1\right\rangle $ transition but
highly detunned (decoupled) from the transitions between other levels of
each qutrit [Fig.~3(c)]. Under this consideration, the interaction
Hamiltonian in the interaction picture and after the RWA can be written as
\begin{equation}
H_{\mathrm{I}_{3}}=\hbar \left( \mu _{l}a_{l}^{+}\left\vert 0\right\rangle
_{l}\left\langle 1\right\vert +\text{h.c.}\right) +\hbar \left(
g_{l}a_{l}^{+}\left\vert 0\right\rangle _{a}\left\langle 1\right\vert +\text{%
h.c.}\right) ,
\end{equation}%
where $\mu _{l}$ ($g_{l}$) is the coupling constant between the cavity mode
and the $\left\vert 0\right\rangle \leftrightarrow \left\vert 1\right\rangle
$ transition of qutrit $l$ (qutrit $a$).

Assume $\mu _{l}=g_{l}.$ This condition can be readily satisfied, because
the coupling strength $\mu _{l}$ can be adjusted by varying the position of
qutrit $l$ in cavity $l$. For solid-state qutrits (e.g., superconducting
qutrits or quantum dots), the condition can also be met by adjusting the
coupling strength $g_{l}$ through varying the coupling element (e.g.,
capacitance or inductance) between qutrit $a$ and cavity $l$. Thus, under
the Hamiltonian (5), one can obtain the following state evolution:
\begin{eqnarray}
\left\vert 0\right\rangle _{l}\left\vert 0\right\rangle _{a}\left\vert
0\right\rangle _{c_{l}} &\rightarrow &\left\vert 0\right\rangle
_{l}\left\vert 0\right\rangle _{a}\left\vert 0\right\rangle _{c_{l}},  \notag
\\
\left\vert 0\right\rangle _{l}\left\vert 1\right\rangle _{a}\left\vert
0\right\rangle _{c_{l}} &\rightarrow &\frac{1}{2}\left( 1+\cos \sqrt{2}%
g_{l}t\right) \left\vert 0\right\rangle _{l}\left\vert 1\right\rangle
_{a}\left\vert 0\right\rangle _{c_{l}}-i\frac{\sqrt{2}}{2}\sin \sqrt{2}%
g_{l}t\left\vert 0\right\rangle _{l}\left\vert 0\right\rangle _{a}\left\vert
1\right\rangle _{c_{l}}  \notag \\
&&-\frac{1}{2}\left( 1-\cos \sqrt{2}g_{l}t\right) \left\vert 1\right\rangle
_{l}\left\vert 0\right\rangle _{a}\left\vert 0\right\rangle _{c_{l}}.
\end{eqnarray}

The results (2), (4), and (6) presented above will be employed for the gate
realization, as shown in the next section.

\begin{center}
\textbf{III. MULTI-QUBIT CONTROLLED PHASE GATE }
\end{center}

To begin with, it should be mentioned that the two logic states 0 and 1 of a
qubit are represented by the two lowest levels $\left\vert0\right\rangle$
and $\left\vert 1\right\rangle $ of a qutrit. Throughout this paper, qubit $%
l $ in cavity $l$ corresponds to qutrit $l$ placed in cavity $l$ ($%
l=1,2,...,n$). Namely, the $n$ qubits ($1,2,...,n$) below, which are
distributed in $n$ different cavities, correspond to the $n$ qutrits ($%
1,2,...,n$) placed in $n$ different cavities, respectively. In this section,
we will first show how to implement a three-qubit controlled phase gate with
three qubits distributed in three different cavities, and then give a
discussion on the realization of an $n$-qubit controlled phase gate with $n$
qubits distributed in $n$ different cavities.

\begin{figure}[tbp]
\begin{center}
\includegraphics[bb=0 0 1700 800, width=14.0 cm, clip]{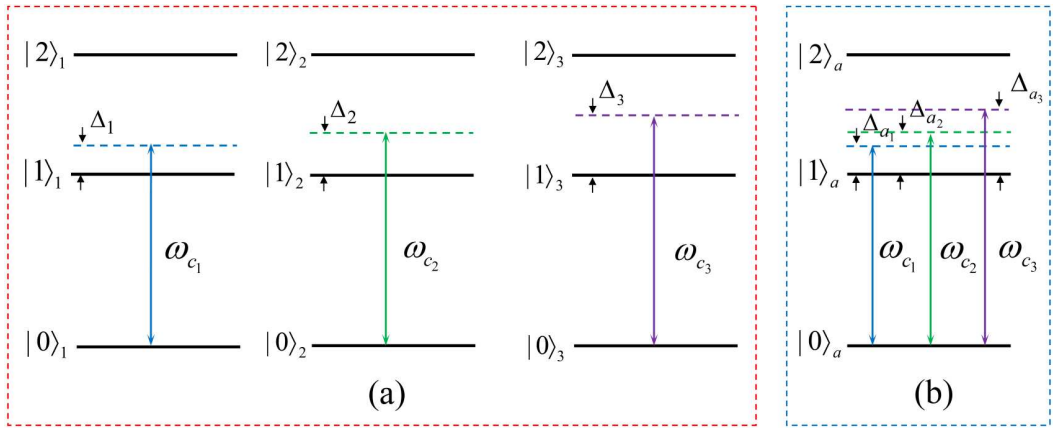} \vspace*{%
-0.08in}
\end{center}
\caption{(Color online) (a) Decoupling of qutrits ($1,2,3$) from their
respective cavities before the gate operation (from left to right). Here, $%
\Delta_{l}$ is a large detuning between the frequency $\protect\omega_{c_l}$
of cavity $l$ and the $\left\vert 0\right\rangle \leftrightarrow \left\vert
1\right\rangle $ transition frequency of qutrit $l$ ($l=1,2,3$), which
represents that cavity $l$ is far-off resonant with (decoupled from) the $%
\left\vert 0\right\rangle \leftrightarrow \left\vert 1\right\rangle $
transition of qutrit $l$. (b) Decoupling of qutrit $a$ from three cavities
before the gate operation. Here, $\Delta_{a_{l}}$ is a large detuning
between the frequency $\protect\omega_{c_l}$ of cavity $l$ and the $%
\left\vert 0\right\rangle \leftrightarrow \left\vert 1\right\rangle $
transition frequency of qutrit $a$ ($l=1,2,3$), which represents that cavity
$l$ is far-off resonant with (decoupled from) $\left\vert 0\right\rangle
\leftrightarrow \left\vert 1\right\rangle $ transition of qutrit $a$.
Because the $\left\vert 1\right\rangle \leftrightarrow \left\vert
2\right\rangle $ level spacing is smaller than the $\left\vert
0\right\rangle \leftrightarrow \left\vert 1\right\rangle $ level spacing,
there is a \textit{larger} detuning between the frequency of cavity $l$ and
the $\left\vert 1\right\rangle \leftrightarrow \left\vert 2\right\rangle $
transition frequency of qutrits $l$ and $a$. Hence, cavity $l$ is
automatically far-off resonant with (or decoupled from) the $\left\vert
1\right\rangle \leftrightarrow \left\vert 2\right\rangle $ transition of
qutrits $l$ and $a$. In addition, the coupling of cavity $l$ with the $%
\left\vert 0\right\rangle \leftrightarrow \left\vert 2\right\rangle $
transition of qutrits $l$ and $a$ is negligible because the $\left\vert
0\right\rangle \leftrightarrow \left\vert 2\right\rangle $ level spacing is
much greater than the $\left\vert 0\right\rangle \leftrightarrow \left\vert
1\right\rangle $ level spacing. For simplicity, we consider each qutrit is
identical, resulting in $\Delta_{l}=\Delta_{a_{l}}$. During the gate
operation, we need to bring the $\left\vert 0\right\rangle \leftrightarrow
\left\vert 1\right\rangle $ transition of qutrit $a$ on resonance with
cavity $l$, corresponding to $\Delta_{a_{l}}=0$ ($l=1,2,3$). In addition, we
need to bring the $\left\vert 0\right\rangle \leftrightarrow \left\vert
1\right\rangle $ transition of both qutrit $l$ and qutrit $a$ on resonance
with cavity $l$, corresponding to $\Delta_{l}=\Delta_{a_{l}}=0$ ($l=1,2,3$).}
\label{fig:4}
\end{figure}

\begin{center}
\textbf{A. Implementing a three-qubit controlled phase gate}
\end{center}

For three qubits, there are a total number of eight computational basis
states, denoted by $\left\vert 000\right\rangle ,$ $\left\vert
001\right\rangle ,$\ldots , $\left\vert 111\right\rangle ,$ respectively. A
three-qubit controlled phase gate is described by $\left\vert
000\right\rangle $ $\rightarrow $ $\left\vert 000\right\rangle ,\left\vert
001\right\rangle $ $\rightarrow $ $\left\vert 001\right\rangle ,\left\vert
010\right\rangle $ $\rightarrow $ $\left\vert 010\right\rangle ,\left\vert
011\right\rangle $ $\rightarrow $ $\left\vert 011\right\rangle ,\left\vert
100\right\rangle $ $\rightarrow $ $\left\vert 100\right\rangle ,\left\vert
101\right\rangle $ $\rightarrow $ $\left\vert 101\right\rangle ,\left\vert
110\right\rangle $ $\rightarrow $ $\left\vert 110\right\rangle ,\left\vert
111\right\rangle $ $\rightarrow -$ $\left\vert 111\right\rangle ,$ which
implies that if and only if the two control qubits (the first two qubits)
are in the state $\left\vert 1\right\rangle $, a phase flip happens to the
state $\left\vert 1\right\rangle $ of the target qubit (the last qubit) but
nothing happens otherwise.

Let us return to the setup shown in Fig. 2(a). For the three-qubit case
(i.e., $n=3$), Fig. 2(a) consists three cavities ($1,2,3)$ each hosting a
qutrit and coupled to an auxiliary qutrit $a$. Initially, the three qutrits (%
$1,2,3$) distributed in the three cavities are decoupled from their
respective cavities [Fig.~4(a)], qutrit $a$ is decoupled from all cavities
[Fig.~4(b)], and each cavity is in the vacuum state. The 3-qubit controlled
phase gate described above can be implemented by the following eight steps
of operation, which are described as follows.

Step 1: Apply a classical pulse to each of qutrits ($2,3$), which has an
initial phase $-\pi /2$ and is resonant with the $\left\vert 1\right\rangle
\leftrightarrow \left\vert 2\right\rangle $ transition. The Rabi frequency
of the pulse applied to qutrit $l$ is $\Omega _{l}$ ($l=2,3$). After an
interaction time $\tau _{1}=\pi /\left( 2\Omega _{l}\right) ,$ the state $%
\left\vert 1\right\rangle $ of qutrit $l$ changes to $\left\vert
2\right\rangle $ according to Eq.~(2). Because of the level spacings of each
qutrit being not adjusted (Fig.~4), each qutrit is decoupled from its cavity
during the pulse.

Step 2: Bring the $\left\vert 0\right\rangle \leftrightarrow \left\vert
1\right\rangle $ transition of qutrits $1$ and $a$ to resonance with cavity $%
1$ for an interaction time $\tau _{2}=\pi /\left( \sqrt{2}g_{1}\right) $
(Fig.~4 with $\Delta _{1}=\Delta _{a_{1}}=0$), resulting in the state
transformation $\left\vert 1\right\rangle _{1}\left\vert 0\right\rangle
_{a}\left\vert 0\right\rangle _{c_{1}}\rightarrow -\left\vert 0\right\rangle
_{1}\left\vert 1\right\rangle _{a}\left\vert 0\right\rangle _{c_{1}}$
according to Eq.~(6).

\begin{figure}[tbp]
\begin{center}
\includegraphics[bb=76 10 850 600, width=12.0 cm, clip]{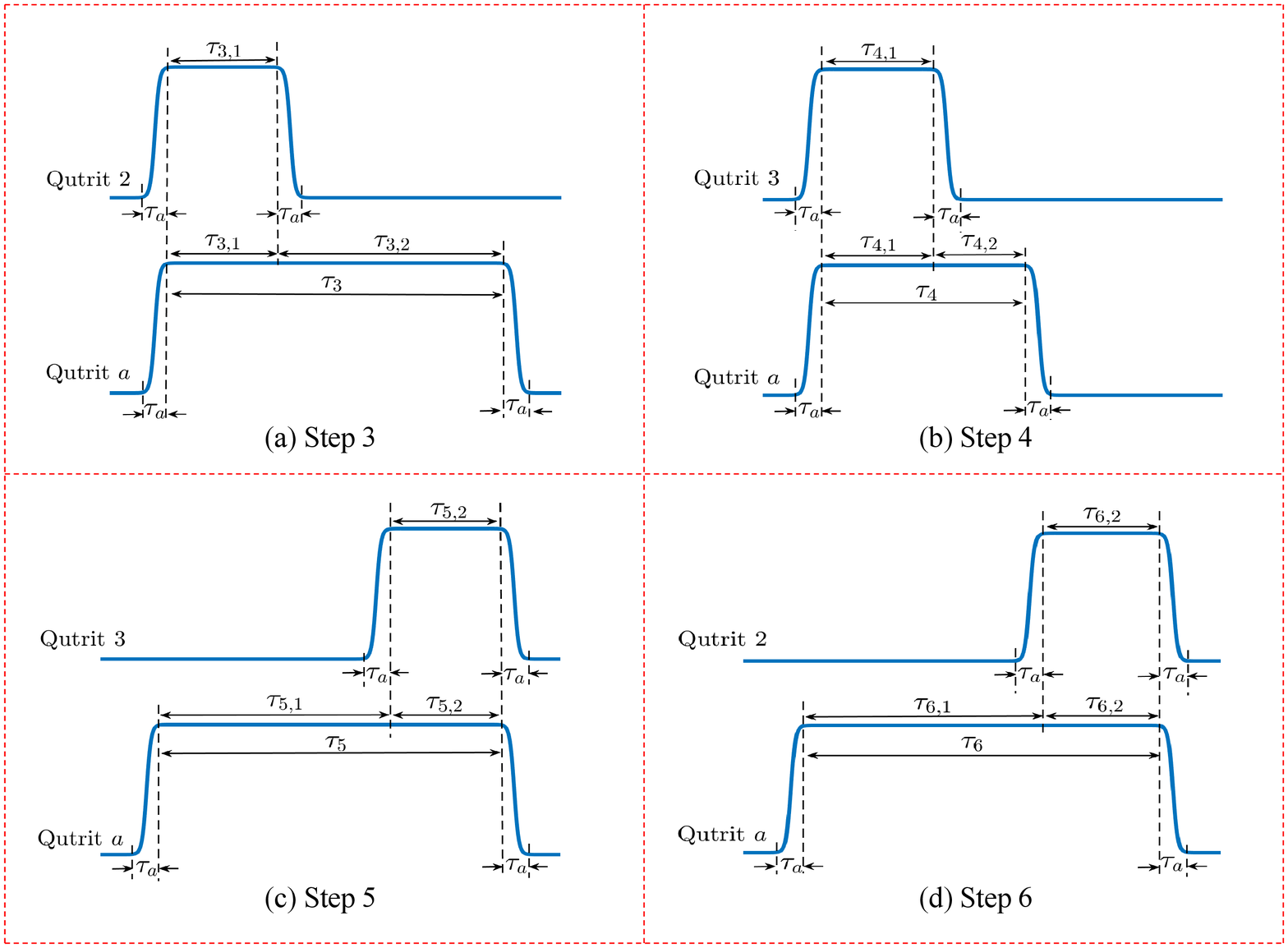} \vspace*{%
-0.08in}
\end{center}
\caption{(Color online) (a) Sequence of operations for step 3. The top
figure corresponds to qutrit 2 while the bottom figure corresponds to qutrit
$a$. (b) Sequence of operations for step 4. The top and bottom figures
correspond to qutrit 3 and qutrit $a$, respectively. (c) Sequence of
operations for step 5. The top figure corresponds to qutrit 3 while the
bottom figure corresponds to qutrit $a$. (d) Sequence of operations for step
6. The top and bottom figures correspond to qutrit 2 and qutrit $a$,
respectively. Here, $\protect\tau _{k,1}, \protect\tau _{k,2}$, and $\protect%
\tau _{k}$ (with $k=3,4,5,6$) are the qutrit-cavity interaction times, as
described in the text; while $\protect\tau _{a}$ is the typical time
required to adjust the qutrit level spacings. Note that the operations
in(a)--(d)are performed from left to right.}
\label{fig:5}
\end{figure}

Step 3: Bring the $\left\vert 0\right\rangle \leftrightarrow \left\vert
1\right\rangle $ transition of qutrits 2 and $a$ to resonance with cavity 2
for an interaction time $\tau _{3,1}=\pi /\left( \sqrt{2}g_{2}\right) $
(Fig.~4 with $\Delta _{2}=\Delta _{a_{2}}=0$)$,$ resulting in $\left\vert
0\right\rangle _{2}\left\vert 1\right\rangle _{a}\left\vert 0\right\rangle
_{c_{2}}\rightarrow -\left\vert 1\right\rangle _{2}\left\vert 0\right\rangle
_{a}\left\vert 0\right\rangle _{c_{2}}$ according to Eq.~(6). Then turn off
the interaction between qutrit 2 and cavity $2$ such that the state $%
-\left\vert 1\right\rangle _{2}\left\vert 0\right\rangle _{a}\left\vert
0\right\rangle _{c_{2}}$ remains unchanged, while let qutrit $a$ continue
resonantly interacting with cavity 2 for an additional interaction time $%
\tau _{3,2}=2\pi /g_{2}-\pi /\left( \sqrt{2}g_{2}\right) $ (Fig.~4 with $%
\Delta _{2}\neq 0$ but $\Delta _{a_{2}}=0$)$.$ After an interaction time $%
\tau _{3}=\tau _{3,1}+$ $\tau _{3,2}=2\pi /g_{2},$ the state $\left\vert
1\right\rangle _{a}\left\vert 0\right\rangle _{c_{2}}$ remains unchanged
according to Eq.~(4). The operation sequence for this step of operation is
illustrated in Fig. 5(a).

Step 4: Bring the $\left\vert 0\right\rangle \leftrightarrow \left\vert
1\right\rangle $ transition of qutrits $3$ and $a$ to resonance with cavity $%
3$ for an interaction time $\tau _{4,1}=\pi /\left( \sqrt{2}g_{3}\right) $
(Fig.~4 with $\Delta _{3}=\Delta _{a_{3}}=0$)$,$ resulting in $\left\vert
0\right\rangle _{3}\left\vert 1\right\rangle _{a}\left\vert 0\right\rangle
_{c_{3}}\rightarrow -\left\vert 1\right\rangle _{3}\left\vert 0\right\rangle
_{a}\left\vert 0\right\rangle _{c_{3}}$ according to Eq. (6). Then, turn off
the interaction between qutrit $3$ and cavity $3$ such that the state $%
-\left\vert 1\right\rangle _{3}\left\vert 0\right\rangle _{a}\left\vert
0\right\rangle _{c_{3}}$ remains unchanged, while let qutrit $a$ continue
resonantly interacting with cavity $3$ for an additional interaction time $%
\tau _{4,2}=\pi /g_{3}-\pi /\left( \sqrt{2}g_{3}\right) $ (Fig.~4 with $%
\Delta _{3}\neq 0$ but $\Delta _{a_{3}}=0$)$.$ After an interaction time $%
\tau _{4}=\tau _{4,1}+$ $\tau _{4,2}=\pi /g_{3}$, the state $\left\vert
1\right\rangle _{a}\left\vert 0\right\rangle _{c_{3}}$ becomes $-\left\vert
1\right\rangle _{a}\left\vert 0\right\rangle _{c_{3}}$ according to Eq.~(4).
The operation sequence for this step of operation is illustrated in Fig.
5(b).

Step 5: Bring the $\left\vert 0\right\rangle \leftrightarrow \left\vert
1\right\rangle $ transition of qutrit $a$ to resonance with cavity 3 for an
interaction time $\tau _{5,1}=2\pi /g_{3}-\pi /\left( \sqrt{2}g_{3}\right) $
(Fig.~4 with $\Delta _{a_{3}}=0$). Then, let qutrit $a$ continue resonantly
interacting with cavity $3$ and bring the $\left\vert 0\right\rangle
\leftrightarrow \left\vert 1\right\rangle $ transition of qutrit $3$ to
resonance with cavity $3$ for an interaction time $\tau _{5,2}=\pi /\left(
\sqrt{2}g_{3}\right) $ (Fig.~4 with $\Delta _{3}=\Delta _{a_{3}}=0$)$.$
After the interaction time $\tau _{5,2},$ we have the state transformation $%
\left\vert 1\right\rangle _{3}\left\vert 0\right\rangle _{a}\left\vert
0\right\rangle _{c_{3}}\rightarrow -\left\vert 0\right\rangle _{3}\left\vert
1\right\rangle _{a}\left\vert 0\right\rangle _{c_{3}}$ according to Eq. (6),
while after an interaction time $\tau _{5}=\tau _{5,1}+\tau _{5,2}=2\pi
/g_{3}$ the state $\left\vert 1\right\rangle _{a}\left\vert 0\right\rangle
_{c_{3}}$ remains unchanged according to Eq. (4). The operation sequence for
this step of operation is illustrated in Fig. 5(c).

Step $6$: Bring the $\left\vert 0\right\rangle \leftrightarrow \left\vert
1\right\rangle $ transition of qutrit $a$ to resonance with cavity $2$ for
an interaction time $\tau _{6,1}=2\pi /g_{2}-\pi /\left( \sqrt{2}%
g_{2}\right) $ (Fig.~4 with $\Delta _{a_{2}}=0$)$.$ Then, let qutrit $a$
continue resonantly interacting with cavity $2$ and bring the $\left\vert
0\right\rangle \leftrightarrow \left\vert 1\right\rangle $ transition of
qutrit $2$ to resonance with cavity $2$ for an interaction time $\tau
_{6,2}=\pi /\left( \sqrt{2}g_{2}\right) $ (Fig.~4 with $\Delta _{2}=\Delta
_{a_{2}}=0$)$.$ After the interaction time $\tau _{6,2},$ one has the state
transformation $\left\vert 1\right\rangle _{2}\left\vert 0\right\rangle
_{a}\left\vert 0\right\rangle _{c_{2}}\rightarrow -\left\vert 0\right\rangle
_{2}\left\vert 1\right\rangle _{a}\left\vert 0\right\rangle _{c_{2}}$
according to Eq. (6), while after an interaction time $\tau _{6}=\tau
_{6,1}+\tau _{6,2}=2\pi /g_{2}$ the state $\left\vert 1\right\rangle
_{a}\left\vert 0\right\rangle _{c_{2}}$ remains unchanged according to Eq.
(4). The operation sequence for this step of operation is illustrated in
Fig. 5(d).

Step $7$: Bring the $\left\vert 0\right\rangle \leftrightarrow \left\vert
1\right\rangle $ transition of qutrits $1$ and $a$ to resonance with cavity $%
1$ for an interaction time $\tau _{7}=\pi /\left( \sqrt{2}g_{1}\right) $
(Fig.~4 with $\Delta _{1}=\Delta _{a_{1}}=0$), resulting in the state
transformation $\left\vert 0\right\rangle _{1}\left\vert 1\right\rangle
_{a}\left\vert 0\right\rangle _{c_{1}}\rightarrow -\left\vert 1\right\rangle
_{1}\left\vert 0\right\rangle _{a}\left\vert 0\right\rangle _{c_{1}}$
according to Eq.~(6).

Step 8$:$ Bring the $\left\vert 0\right\rangle \leftrightarrow \left\vert
1\right\rangle $ transition of each qutrit to off-resonance with its cavity
or cavities (Fig.~4 with non-zero detunings), such that the qutrit system is
decoupled from the cavity system. Then, apply a classical pulse to each of
qutrits ($2,3$), which has an initial phase $\pi /2$ and is resonant with
the $\left\vert 1\right\rangle \leftrightarrow \left\vert 2\right\rangle $
transition of the qutrits. The Rabi frequency of the pulse applied to qutrit
$l$ is $\Omega _{l}$ ($l=2,3$). After a pulse duration $\tau _{8}=\pi
/\left( 2\Omega _{l}\right) ,$ the state $\left\vert 2\right\rangle $ of
qutrit $l$ changes to $\left\vert 1\right\rangle $ according to Eq.~(2).

One can check that the states of the whole system after each step of the
above operation are summarized below:%
\begin{eqnarray}
&&\text{ \ \ \ \ \ \ }%
\begin{array}{c}
\left\vert 100\right\rangle \left\vert 0\right\rangle _{a}\left\vert
0\right\rangle _{c} \\
\left\vert 101\right\rangle \left\vert 0\right\rangle _{a}\left\vert
0\right\rangle _{c} \\
\left\vert 110\right\rangle \left\vert 0\right\rangle _{a}\left\vert
0\right\rangle _{c} \\
\left\vert 111\right\rangle \left\vert 0\right\rangle _{a}\left\vert
0\right\rangle _{c}%
\end{array}%
\overset{\text{Step 1}}{\longrightarrow }%
\begin{array}{c}
\left\vert 100\right\rangle \left\vert 0\right\rangle _{a}\left\vert
0\right\rangle _{c} \\
\left\vert 102\right\rangle \left\vert 0\right\rangle _{a}\left\vert
0\right\rangle _{c} \\
\left\vert 120\right\rangle \left\vert 0\right\rangle _{a}\left\vert
0\right\rangle _{c} \\
\left\vert 122\right\rangle \left\vert 0\right\rangle _{a}\left\vert
0\right\rangle _{c}%
\end{array}%
\overset{\text{Step 2}}{\longrightarrow }%
\begin{array}{c}
-\left\vert 000\right\rangle \left\vert 1\right\rangle _{a}\left\vert
0\right\rangle _{c} \\
-\left\vert 002\right\rangle \left\vert 1\right\rangle _{a}\left\vert
0\right\rangle _{c} \\
-\left\vert 020\right\rangle \left\vert 1\right\rangle _{a}\left\vert
0\right\rangle _{c} \\
-\left\vert 022\right\rangle \left\vert 1\right\rangle _{a}\left\vert
0\right\rangle _{c}%
\end{array}
\notag \\
&&\overset{\text{Step 3}}{\longrightarrow }%
\begin{array}{c}
\left\vert 010\right\rangle \left\vert 0\right\rangle _{a}\left\vert
0\right\rangle _{c} \\
\left\vert 012\right\rangle \left\vert 0\right\rangle _{a}\left\vert
0\right\rangle _{c} \\
-\left\vert 020\right\rangle \left\vert 1\right\rangle _{a}\left\vert
0\right\rangle _{c} \\
-\left\vert 022\right\rangle \left\vert 1\right\rangle _{a}\left\vert
0\right\rangle _{c}%
\end{array}%
\overset{\text{Step 4}}{\longrightarrow }%
\begin{array}{c}
\left\vert 010\right\rangle \left\vert 0\right\rangle _{a}\left\vert
0\right\rangle _{c} \\
\left\vert 012\right\rangle \left\vert 0\right\rangle _{a}\left\vert
0\right\rangle _{c} \\
\left\vert 021\right\rangle \left\vert 0\right\rangle _{a}\left\vert
0\right\rangle _{c} \\
\left\vert 022\right\rangle \left\vert 1\right\rangle _{a}\left\vert
0\right\rangle _{c}%
\end{array}%
\overset{\text{Step 5}}{\longrightarrow }%
\begin{array}{c}
\left\vert 010\right\rangle \left\vert 0\right\rangle _{a}\left\vert
0\right\rangle _{c} \\
\left\vert 012\right\rangle \left\vert 0\right\rangle _{a}\left\vert
0\right\rangle _{c} \\
-\left\vert 020\right\rangle \left\vert 1\right\rangle _{a}\left\vert
0\right\rangle _{c} \\
\left\vert 022\right\rangle \left\vert 1\right\rangle _{a}\left\vert
0\right\rangle _{c}%
\end{array}
\notag \\
&&\overset{\text{Step 6}}{\longrightarrow }%
\begin{array}{c}
-\left\vert 000\right\rangle \left\vert 1\right\rangle _{a}\left\vert
0\right\rangle _{c} \\
-\left\vert 002\right\rangle \left\vert 1\right\rangle _{a}\left\vert
0\right\rangle _{c} \\
-\left\vert 020\right\rangle \left\vert 1\right\rangle _{a}\left\vert
0\right\rangle _{c} \\
\left\vert 022\right\rangle \left\vert 1\right\rangle _{a}\left\vert
0\right\rangle _{c}%
\end{array}%
\overset{\text{Step 7}}{\longrightarrow }%
\begin{array}{c}
\left\vert 100\right\rangle \left\vert 0\right\rangle _{a}\left\vert
0\right\rangle _{c} \\
\left\vert 102\right\rangle \left\vert 0\right\rangle _{a}\left\vert
0\right\rangle _{c} \\
\left\vert 120\right\rangle \left\vert 0\right\rangle _{a}\left\vert
0\right\rangle _{c} \\
-\left\vert 122\right\rangle \left\vert 0\right\rangle _{a}\left\vert
0\right\rangle _{c}%
\end{array}%
\overset{\text{Step 8}}{\longrightarrow }%
\begin{array}{c}
\left\vert 100\right\rangle \left\vert 0\right\rangle _{a}\left\vert
0\right\rangle _{c} \\
\left\vert 101\right\rangle \left\vert 0\right\rangle _{a}\left\vert
0\right\rangle _{c} \\
\left\vert 110\right\rangle \left\vert 0\right\rangle _{a}\left\vert
0\right\rangle _{c} \\
-\left\vert 111\right\rangle \left\vert 0\right\rangle _{a}\left\vert
0\right\rangle _{c}%
\end{array}%
\end{eqnarray}%
where $\left\vert ijk\right\rangle $ is abbreviation of the state $%
\left\vert i\right\rangle _{1}\left\vert j\right\rangle _{2}\left\vert
k\right\rangle _{3}$ of qubits ($1,2,3$) with $i,j,k\in \{0,1,2\}\ $while $%
\left\vert 0\right\rangle _{c}$ is abbreviation of the state $\left\vert
0\right\rangle _{c_{1}}\left\vert 0\right\rangle _{c_{2}}\left\vert
0\right\rangle _{c_{3}}$ of cavities ($1,2,3$).

On the other hand, it is obvious that the following states of the system
\begin{equation}
\;\left\vert 000\right\rangle \left\vert 0\right\rangle _{a}\left\vert
0\right\rangle _{c},\left\vert 001\right\rangle \left\vert 0\right\rangle
_{a}\left\vert 0\right\rangle _{c},\left\vert 010\right\rangle \left\vert
0\right\rangle _{a}\left\vert 0\right\rangle _{c},\left\vert
011\right\rangle \left\vert 0\right\rangle _{a}\left\vert 0\right\rangle _{c}
\end{equation}%
remain unchanged during the entire operation. This is because: the state $%
\left\vert 1\right\rangle $ of each of qutrits $2$ and $3$ changes to the
state $\left\vert 2\right\rangle $ after applying the pulses (step 1); no
photon was emitted to the cavities during each step of operations above,
when each qutrit is in the state $\left\vert 0\right\rangle $ or $\left\vert
2\right\rangle $ (steps 2-7); and the state $\left\vert 2\right\rangle $ of
each of qutrits $2$ and $3$ changes back to the state $\left\vert
1\right\rangle $ after applying the pulses (step 8). Thus, we can conclude
from Eq.~(7) that the three-qubit controlled phase gate was realized with
two controlled qubits ($1,2$) distributed in two different cavities ($1,2$),
as well as the target qubit $3$ in cavity 3 after the above process.

\begin{center}
\textbf{B. Implementing an }$n$\textbf{-qubit controlled phase gate}
\end{center}

\begin{figure}[tbp]
\begin{center}
\includegraphics[bb=88 176 444 327, width=15.0 cm, clip]{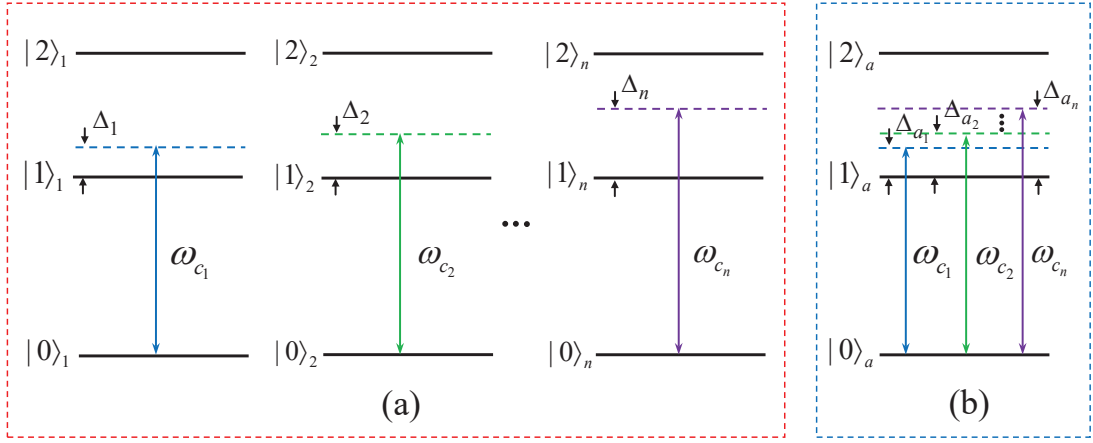} \vspace*{%
-0.08in}
\end{center}
\caption{(Color online) (a) Decoupling of qutrits ($1,2,...,n$) from their
respective cavities before the gate operation (from left to right). Here, $%
\Delta _{l}$ is a large detuning between the frequency $\protect\omega %
_{c_{l}}$ of cavity $l$ and the $\left\vert 0\right\rangle \leftrightarrow
\left\vert 1\right\rangle $ transition frequency of qutrit $l$ ($l=1,2,...,n$%
), representing that cavity $l$ is far-off resonant with (decoupled from)
the $\left\vert 0\right\rangle \leftrightarrow \left\vert 1\right\rangle $
transition of qutrit $l$. (b) Decoupling of qutrit $a$ from all $n$ cavities
before the gate operation. Here, $\Delta _{a_{l}}$ is a large detuning
between the frequency $\protect\omega _{c_{l}}$ of cavity $l$ and the $%
\left\vert 0\right\rangle \leftrightarrow \left\vert 1\right\rangle $
transition frequency of qutrit $a$ ($l=1,2,...,n$), indicating that cavity $%
l $ is far-off resonant with (decoupled from) the $\left\vert 0\right\rangle
\leftrightarrow \left\vert 1\right\rangle $ transition of qutrit $a$. During
the gate operation, one needs to bring the $\left\vert 0\right\rangle
\leftrightarrow \left\vert 1\right\rangle $ transition of qutrit $a$ on
resonance with cavity $l$, which can be met by setting $\Delta _{a_{l}}=0$ ($%
l=1,2,...,n$). In addition, one needs to bring the $\left\vert
0\right\rangle \leftrightarrow \left\vert 1\right\rangle $ transition of
qutrits $l$ and $a$ on resonance with cavity $l$, corresponding to $\Delta
_{l}=\Delta _{a_{l}}=0$ ($l=1,2,...,n$).}
\label{fig:6}
\end{figure}

A $n$-qubit controlled phase gate of $n$ qubits ($1,2,...,n $) is defined by
the following transformation
\begin{equation}
\left\vert i_{1}i_{2}...i_{n}\right\rangle \rightarrow \left( -1\right)
^{i_{1}\times i_{2}...\times i_{n}}\left\vert
i_{1}i_{2}...i_{l}...i_{n}\right\rangle ,
\end{equation}%
where subscript $l$ indicates qubit $l$, $i_{l}\in \left\{ 0,1\right\} $,
and $\left\vert i_{1}i_{2}...i_{l}...i_{n}\right\rangle $ is a $n$-qubit
computational basis state. For $n$ qubits, there exist $2^{n}$ computational
basis states, forming a set of complete orthogonal bases in a $2^{n}$%
-dimensional Hilbert space of the $n$ qubits. Eq.~(9) implies that only when
the $n-1$ control qubits (the first $n-1$ qubits) are all in the state $%
\left\vert 1\right\rangle ,$ the state $\left\vert 1\right\rangle $ of the
target qubit (the last qubit) undergoes a phase flip, i.e., $\left\vert
11...1\right\rangle \rightarrow -\left\vert 11...1\right\rangle ,$ while
nothing happens to all other $2^{n}-1$ computational basis states. In the
following, we will discuss how this multi-qubit controlled phase gate can be
achieved with $n$ qutrits distributed in different cavities.

Let us now consider a setup shown in Fig. 2(a), where each cavity hosts a
qutrit and coupled to an auxiliary qutrit $a.$ Suppose that qutrits ($%
1,2,...,n$), distributed in the $n$ cavities, are initially decoupled from
their respective cavities and qutrit $a$ is decoupled from all cavities
(Fig.~6). Each cavity is initially in the vacuum state. Examining the above
operations for the three-qubit controlled phase gate carefully, we find that
the $n$-qubit controlled phase gate (9) can be obtained with each cavity
returning to the original vacuum state, by the following sequence of
operators
\begin{equation}
U=U_{2n+2}\otimes U_{2n+1}\otimes \left( \prod_{l=1}^{n-1}U_{n+l+1}\right)
\otimes U_{n+1}\otimes \left( \prod_{l=2}^{n-1}U_{l+1}\right) \otimes
U_{2}\otimes U_{1},
\end{equation}%
where $U_{1},U_{2},U_{l+1},U_{n+1},U_{n+l+1},U_{2n+1},$and $U_{2n+2}$
represent unitary operators. The state transformations arising from the
unitary operators are described below:

(i) $U_{1}$ transforms the state $\left\vert 1\right\rangle $ of qutrit $l$
to $\left\vert 2\right\rangle $ ($l=2,3,...,n$). This transformation can be
obtained by: Apply a classical pulse to each of qutrits ($2,3,...,n$), which
is resonant with the $\left\vert 1\right\rangle \leftrightarrow \left\vert
2\right\rangle $ transition of the qutrits. Each pulse has an initial phase $%
-\pi /2$. The duration of the pulse applied to qutrit $l$ is $\pi /\left(
2\Omega _{l}\right) .$

(ii) $U_{2}$ leads to the state transform $\left\vert 1\right\rangle
_{1}\left\vert 0\right\rangle _{a}\left\vert 0\right\rangle
_{c_{1}}\rightarrow -\left\vert 0\right\rangle _{1}\left\vert 1\right\rangle
_{a}\left\vert 0\right\rangle _{c_{1}},$ which can be realized by: Bring the
$\left\vert 0\right\rangle \leftrightarrow \left\vert 1\right\rangle $
transition of qutrits $1$ and $a$ to resonance with cavity $1$ for an
interaction time $\pi /\left( \sqrt{2}g_{1}\right) $.

(iii) $U_{l+1}$ $(l=2,3,\ldots ,n-1)$ results in the state transformation $%
\left\vert 0\right\rangle _{l}\left\vert 1\right\rangle _{a}\left\vert
0\right\rangle _{c_{l}}\rightarrow -\left\vert 1\right\rangle _{l}\left\vert
0\right\rangle _{a}\left\vert 0\right\rangle _{c_{l}}$ while nothing to the
state $\left\vert 1\right\rangle _{a}\left\vert 0\right\rangle _{c_{l}}.$
This can be achieved by: Bring the $\left\vert 0\right\rangle
\leftrightarrow \left\vert 1\right\rangle $ transition of qutrits $l$ and $a$
to resonance with cavity $l$ for an interaction time $\pi /\left( \sqrt{2}%
g_{l}\right) ,$ then turn off the interaction between qutrit $l$ and cavity $%
l$ but let qutrit $a$ continue resonantly interacting with cavity $l$ for an
additional interaction time $2\pi /g_{l}-\pi /\left( \sqrt{2}g_{l}\right) .$

(iv) $U_{n+1}$ leads to the state transformation $\left\vert 0\right\rangle
_{n}\left\vert 1\right\rangle _{a}\left\vert 0\right\rangle
_{c_{n}}\rightarrow -\left\vert 1\right\rangle _{n}\left\vert 0\right\rangle
_{a}\left\vert 0\right\rangle _{c_{n}}$ and $\left\vert 1\right\rangle
_{a}\left\vert 0\right\rangle _{c_{n}}\rightarrow $ $-\left\vert
1\right\rangle _{a}\left\vert 0\right\rangle _{c_{n}}.$ This state
transformation can be realized by: Bring the $\left\vert 0\right\rangle
\leftrightarrow \left\vert 1\right\rangle $ transition of qutrits $n$ and $a$
to resonance with cavity $n$ for an interaction time $\pi /\left( \sqrt{2}%
g_{n}\right) ,$ then turn off the interaction between qutrit $n$ and cavity $%
n$ but let qutrit $a$ continue resonantly interacting with cavity $n$ for an
additional interaction time $\pi /g_{n}-\pi /\left( \sqrt{2}g_{n}\right) .$

(v) $U_{n+l+1}$ $(l=1,2,\ldots ,n-1)$ leads to the state transformation $%
\left\vert 1\right\rangle _{n-l+1}\left\vert 0\right\rangle _{a}\left\vert
0\right\rangle _{c_{n-l+1}}\rightarrow -\left\vert 0\right\rangle
_{n-l+1}\left\vert 1\right\rangle _{a}\left\vert 0\right\rangle _{c_{n-l+1}}$
while nothing to the state $\left\vert 1\right\rangle _{a}\left\vert
0\right\rangle _{c_{n-l+1}}.$ This can be obtained by: Bring the $\left\vert
0\right\rangle \leftrightarrow \left\vert 1\right\rangle $ transition of
qutrit $a$ to resonance with cavity $n-l+1$ for an interaction time $2\pi
/g_{n-l+1}-\pi /\left( \sqrt{2}g_{n-l+1}\right) ,$ then let qutrit $a$
continue resonantly interacting with cavity $n-l+1$ and bring the $%
\left\vert 0\right\rangle \leftrightarrow \left\vert 1\right\rangle $
transition of qutrit $n-l+1$ to resonance with cavity $n-l+1$ for an
interaction time $\pi /\left( \sqrt{2}g_{n-l+1}\right) .$

(vi) $U_{2n+1}$ results in the state transformation $\left\vert
0\right\rangle _{1}\left\vert 1\right\rangle _{a}\left\vert 0\right\rangle
_{c_{1}}\rightarrow -\left\vert 1\right\rangle _{1}\left\vert 0\right\rangle
_{a}\left\vert 0\right\rangle _{c_{1}},$ which can be realized by: Bring the
$\left\vert 0\right\rangle \leftrightarrow \left\vert 1\right\rangle $
transition of qutrits $1$ and $a$ to resonance with cavity $1$ for an
interaction time $\pi /\left( \sqrt{2}g_{1}\right) $.

(vii) $U_{2n+2}$ transforms the state $\left\vert 2\right\rangle $ of qutrit
$l$ ($l=2,3,...,n$) to $\left\vert 1\right\rangle $. This state
transformation can be implemented by: Apply a classical pulse to each of
qutrits ($2,3,...,n$), which is resonant with the $\left\vert 1\right\rangle
\leftrightarrow \left\vert 2\right\rangle $ transition of the qutrits. Each
pulse has an initial phase $\pi /2$. The duration of the pulse applied to
qutrit $l$ is $\pi /\left( 2\Omega _{l}\right) .$

One can check that after the above $2n+2$ basic unitary operations, the $n$%
-qubit controlled phase gate, described by Eq.~(9), was implemented with $%
n-1 $ control qubits (i.e., qubits $1,$ $2,$ $...,$ and $n-1$, respectively
distributed in cavities $1,$ $2,...,$ and $n-1$) and a target qubit (i.e.
qubit $n$ in cavity $n$).

Before ending this section, several points need to be addressed as follows:

(a). The qutrits, not involved in each step of operation, need to be
decoupled from the cavities during the cavity-qutrit interaction. The
cavities need to be not excited during the application of the pulse. In
addition, the unwanted coupling of the levels $\left\vert 0\right\rangle $, $%
\left\vert 1\right\rangle ,$ and $\left\vert 2\right\rangle $, induced by
the cavity or the pulse, should be negligible for each qutrit. In principle,
these conditions can be satisfied by adjusting the level spacings of the
qutrits [34-40]. For example, the level spacings of superconducting qutrits
can be rapidly adjusted by varying external control parameters (e.g., the
magnetic flux applied to a superconducting loop of phase, transmon, Xmon, or
flux qutrits; see, e.g., [34-37]); the level spacings of NV centers can be
readily adjusted by changing the external magnetic field applied along the
crystalline axis of each NV center [38,39]; and the level spacings of
atoms/quantum dots can be adjusted by changing the voltage on the electrodes
around each atom/quantum dot [40].

(b). As shown above, the ancillary qubit does not need to interact all
cavities simultaneously. Instead, the ancillary qubit is only required to
resonantly interact with each cavity one by one, which corresponds to the case when the detunings $%
\Delta _{a_{1}},$ $\Delta _{a_{2}},...,$ and $\Delta _{a_{n}}$ are set to be zero
one by one [Fig. 6(b)]. Note that the detuning $\Delta _{a_{l}}$ ($l=1,2,...,n$) can be set zero, by
adjusting the level spacings of the ancillary qubit or adjusting the
frequency of cavity $l$ such that the transition frequency between the two levels
$\left\vert 0\right\rangle $ and $\left\vert 1\right\rangle $ is
equal to the frequency of cavity $l$. In practice, setting $%
\Delta _{a_{1}},$ $\Delta _{a_{2}},...,$ and $\Delta _{a_{n}}$ to be zero
one by one becomes a challenge in experiments as the number of cavities increases. Note that frequencies of the
microwave cavities or resonators can be quickly tuned in 1--3 ns [41,42].

(c). It is preferable to use single-mode cavities, which can be designed
with appropriate choice of cavity parameters. However, using single-mode
cavities is not necessary. Multi-mode cavities may be used because one can
choose one mode to interact with the qutrits while have all other modes well
decoupled from the three levels of the qutrits. This can be achieved by
choosing qutrits with a proper level structure or designing the level
structure of qutrits (e.g., solid-state qutrits) with a proper choice of
device parameters. In addition, the method presented here is applicable to
1D, 2D, or 3D cavities or resonators as long as the conditions described
above can be met.

(d). A multi-qubit controlled NOT (CNOT) gate, with multiple qubits
simultaneously controlling a single target qubit, is often called as a
Toffoli gate, for which the state $\left\vert 0\right\rangle $ ($\left\vert
1\right\rangle $) of the target qubit flips to $\left\vert 1\right\rangle $ (%
$\left\vert 0\right\rangle $) when all control qubits are in the state $%
\left\vert 11...1\right\rangle $ but nothing happens to the state of the
target qubit otherwise. As is well known, an $n$-qubit Toffoli gate can be
constructed by using the $n$-qubit controlled phase gate (9) plus two
single-qubit Hadamard gates, which are performed on the target qubit before
and after the $n$-qubit controlled phase gate (9) respectively. Each of the
single-qubit Hadamard transformations, $\left\vert 0\right\rangle
\rightarrow \left( \left\vert 0\right\rangle +\left\vert 1\right\rangle
\right) /\sqrt{2}$ and $\left\vert 1\right\rangle \rightarrow \left(
\left\vert 0\right\rangle -\left\vert 1\right\rangle \right) /\sqrt{2},$ can
be performed by applying a $\pi /2$ classical pulse resonant with the $%
\left\vert 0\right\rangle \leftrightarrow \left\vert 1\right\rangle $
transition of the target qubit. By combining with the above controlled-phase
gate operations, one can implement an $n$-qubit Toffoli gate with only $2n+4$
basic operations. Thus, the present method also provides a simple way to
realize a multi-qubit Toffoli gate with qubits distributed in different
cavities.

\begin{center}
\textbf{IV. DISCUSSION}
\end{center}

In this section we discuss issues that are important to experimental
implementation. For the method to work, the following requirements need to
be satisfied. First, the total operation time
\begin{equation}
\tau =\pi /\Omega +\sqrt{2}\pi /g_{1}+\sum\limits_{j=3}^{n}4\pi
/g_{j-1}+3\pi /g_{n}+\left( 6n-5\right) \tau _{a}
\end{equation}%
should be much shorter than the energy relaxation time $T_{1}$ and dephasing
time $T_{2}$ of the level $\left\vert 2\right\rangle $ of qutrits. Here, we
have set $\Omega _{l}\equiv \Omega $ for simplicity, which can be achieved
by adjusting the intensity of pluses applied to the qutrits, and $\tau _{a}$
is the typical time required for adjusting the level spacings of a single
qutrit. Second, the lifetime of the mode of each cavity is given by $\kappa
_{l}^{-1}=Q_{l}/\omega _{c_{l}}$ with $Q_{l}$ being the quality factor of
cavity $l$ ($l=1,2,...,n$), which should be much longer than a single
qutrit-cavity interaction time. Last, crosstalk between different cavities
needs to be negligible since this interaction is not intended.

These requirements can in principle be realized, since (i) $\tau $ can be
reduced by increasing the coupling constant $g_{l}$ ($l=1,2,...,n$) (i.e.,
each qutrit is placed at an antinode of the cavity mode), (ii) $\tau _{a}$
can be shortened by rapid adjustment of the level spacings of the qutrits
(e.g., the typical time is $1-3$ ns for adjusting the level spacings of a
superconducting qutrit [35,43,44]), (iii) $\kappa _{l}^{-1}$ can be
increased by employing a high-$Q_{l}$ cavity so that the cavity dissipation
is negligible during the operation, (iv) the qutrits can be chosen or
designed so that the energy relaxation time $T_{1}$ and the dephasing time $%
T_{2}$ of the level $\left\vert 2\right\rangle $ are sufficiently long, and
(v) direct crosstalk between cavities can be made negligibly small as long
as $\Delta _{kl}>>g_{kl}$. Here, $\Delta _{kl}$ is the frequency difference
between cavities $k$ and $l$, $g_{kl}$ is the coupling strength between
cavities $k$ and $l$, with $k,l\in \{1,2,...,n\}$ and $k\neq l$.

\begin{figure}[tbp]
\begin{center}
\includegraphics[bb=63 304 358 596, width=10.5 cm, clip]{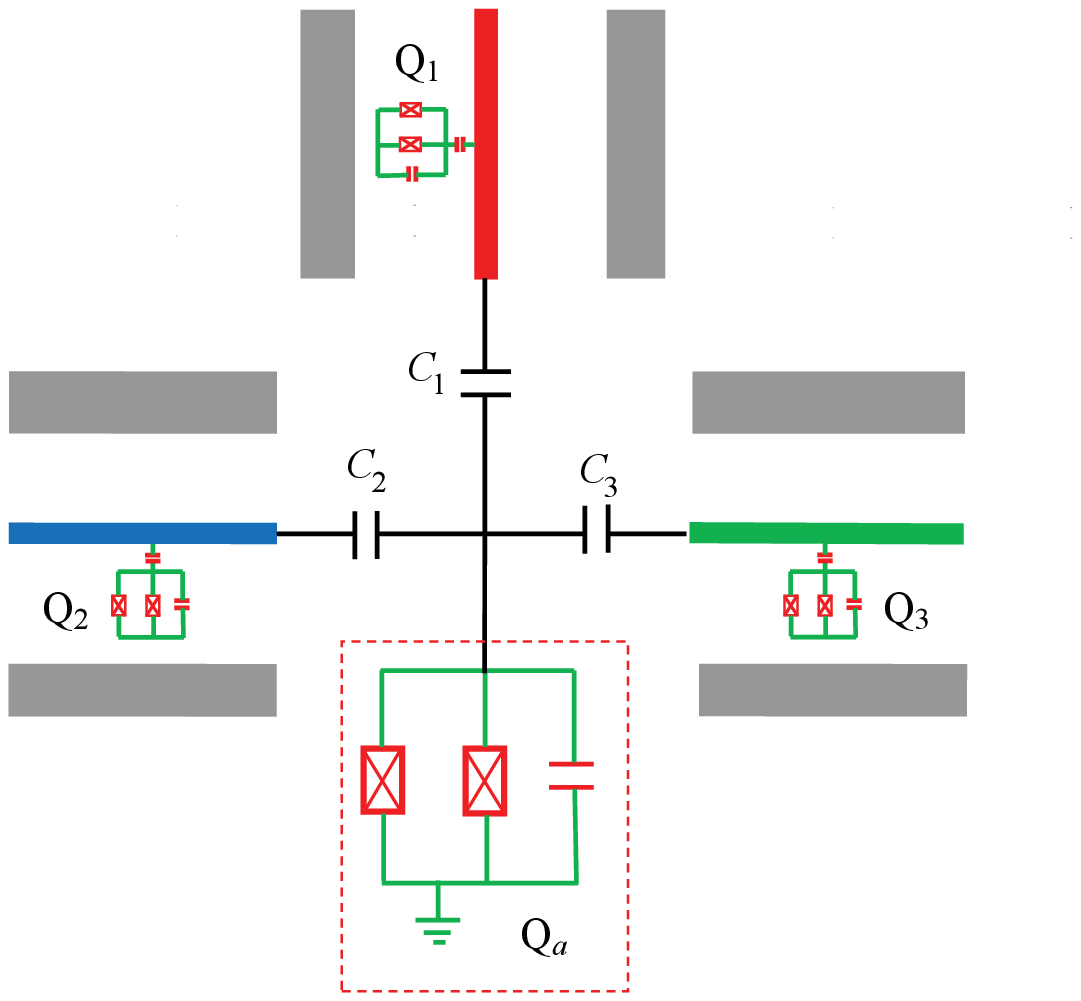} \vspace*{%
-0.08in}
\end{center}
\caption{(Color online) Setup for three cavities each hosting a superconducting
transmon qutrit and coupled by an ancillary transmon qutrit. Each cavity here is a
one-dimensional coplanar resonator, consisting of a central waveguide line and two
lateral ground planes. $Q_1, Q_2,$ and $Q_3$ represent three transmon qutrits 1, 2, and 3,
which are capacitively coupled to their respective cavities. $Q_a$ represents the coupler
qutrit $a$, which is capacitively coupled to the three cavities via capacitance $C_1$, $C_2$,
and $C_3$, respectively. The electronic circuit of a transmon qutrit consists of two Josephson
junctions and a capacitor.}
\label{fig:7}
\end{figure}

\begin{center}
\textbf{V. POSSIBLE EXPERIMENTAL IMPLEMENTATION IN CIRCUIT QED}
\end{center}

Circuit QED, consisting of microwave cavities and superconducting qubits,
is particularly attractive and considered as one of the leading
candidates for QIP [45-52]. In the past decade, based on circuit QED, much progress has been made
in quantum state preparation, quantum logic gate implementation,  quantum state transfer, etc.
with superconducting qubits or microwave photons (see the review articles [49-52]).
For the sake of definitiveness, let us now consider the experimental feasibility of realizing a three-qubit controlled
phase gate, which act on three qubits distributed in three different
cavities. The three-qubit gate here is implemented by using a circuit QED
system, which consists of three identical superconducting transmon qutrits ($%
1,2,3$) respectively embedded in three cavities ($1,2,3$), which are coupled
to an auxiliary transmon qutrit $a$ (Fig.~7). Note that the three
qubits (1,2,3) involved in the gate correspond to the three transmon qutrits
(1,2,3), respectively. For a transmon qutrit, its design is closely related to the Cooper pair box (CPB) with two Josephson junctions [63]. Thus, it is noted that the coupling mechanism of a transmon qutrit and a cavity is the same as that of a CPB and a cavity.

Each cavity considered in Fig.7 is a one-dimensional coplanar resonator (TLR),
which consists of a central waveguide line and two lateral ground planes. After quantization of the cavity field,
there is a standing-wave distribution of quantum voltage along the central waveguide line of the cavity.
The quantum voltage is caused by the electric field of the cavity mode. As shown in Fig. 7, the transmon qutrit hosted in each cavity is capacitively coupled to the cavity, by a capacitor connecting to the central waveguide line of the TLR and the transmon qutrit [37,63,53-55]. This capacitive coupling is induced due to the quantum voltage exerted on the coupling capacitor [63] (also, see [4,47]).
Alternatively, the transmon qutrit hosted in each cavity can be coupled to the cavity by the cavity magnetic field threading the superconducting loop of the qutrit [56-58]. The qutrit-cavity coupling strength can be varied by adjusting the loop size of the qutrit, the
position of the qutrit in the cavity, or the coupling capacitance.

In Fig. 7, the coupling mechanism for the ancillary transmon qutrit capacitively coupled
to each cavity is the same as that for the intra-cavity transmon qutrit capacitively coupled
to the corresponding cavity. To have the ancillary transmon qutrit coupled to cavity $l$ ($l=1,2,3$),
the quantum voltage at one end of the central guide line of cavity $l$, which connects
the coupling capacitor $C_l$, should be not zero. This requirement can be met by a prior
design of cavity, such that that end of the central waveguide line is an antinode of the quantum voltage.
It is noted that the coupling strength of the ancillary transmon qutrit with
each cavity can be varied by a prior design of the sample with an appropriate coupling
capacitance.

From the description given in Sec. III, the gate implementation involves the
following three basic operations:

(i) The first basic operation is described by the Hamiltonian $H_{\mathrm{I}%
_{1}}$ in Eq. (1). In reality, the inter-cavity crosstalk between the two
cavities is inevitable [59], and there exists the unwanted coupling of the
pulse with the $|0\rangle \leftrightarrow |1\rangle $ transition. When these
factors are taken into account, the Hamiltonian $H_{\mathrm{I}_{1}}$ is
modified as
\begin{equation}
\widetilde{H}_{\mathrm{I}_{1}}=\hbar \left( \Omega _{l}e^{i\phi
_{l}}\left\vert 1\right\rangle _{l}\left\langle 2\right\vert +\text{h.c.}%
\right) +\hbar \left( \widetilde{\Omega }_{l}e^{i\phi _{l}}e^{-i\widetilde{%
\delta }_{l}t}\left\vert 0\right\rangle _{l}\left\langle 1\right\vert +\text{%
h.c.}\right) +\varepsilon ,
\end{equation}%
where the first bracket term represents the resonant interaction of the
pulse with the $|1\rangle \leftrightarrow |2\rangle $ transition of qutrit $l
$ ($l=2,3$), while the second bracket term represents the unwanted
interaction of the pulse with the $|0\rangle \leftrightarrow |1\rangle $
transition with Rabi frequency $\widetilde{\Omega }_{l}$ and detuning $%
\widetilde{\delta }_{l}=\widetilde{\omega }_{10,l}-\omega _{p,l}>0$ ($\omega
_{p,l}$ is the frequency of the pulse applied to qutrit $l$) [Fig.~8(a)].
Note that the transition frequencies of qutrit $l$ when interacting with a
classical pulse are different from those of qutrit $l$ when interacting with
cavity $l$. Thus, we here define $\widetilde{\omega }_{10,l}$ ($\widetilde{%
\omega }_{21,l}$) as the $|0\rangle \leftrightarrow |1\rangle $ ($|1\rangle
\leftrightarrow |2\rangle $) transition frequency of qutrit $l$ when
interacting with a pulse, but we will later use different notations to
define the transition frequencies of qutrit $l$ when interacting with cavity
$l$. The detuning $\widetilde{\delta }_{l}$ can be further written as $%
\widetilde{\delta }_{l}=\widetilde{\omega }_{10,l}-\widetilde{\omega }%
_{21,l}=\delta \widetilde{\omega _{l}}$ because of $\omega _{p,l}=\widetilde{%
\omega }_{21,l}$. Here, $\delta \widetilde{\omega }_{l}$ is the
anharmonicity between the $|0\rangle \leftrightarrow |1\rangle $ level
spacing and the $|1\rangle \leftrightarrow |2\rangle $ level spacing of
qutrit $l$. The last term $\varepsilon $ in Eq.~(11) represents the
inter-cavity crosstalk of cavities, which is given by

\begin{equation}
\varepsilon =g_{12}\left( e^{i\Delta _{12}t}a_{1}a_{2}^{+}+h.c.\right)
+g_{23}\left( e^{i\Delta _{23}t}a_{2}a_{3}^{+}+h.c.\right) +g_{13}\left(
e^{i\Delta _{13}t}a_{1}a_{3}^{+}+h.c.\right) ,
\end{equation}%
where $g_{12},$ $g_{23},$ and $g_{13}$ are the coupling strengths of
cavities ($1,2$), ($2,3$), and ($1,3$), respectively; while $\Delta
_{12}=\omega _{c_{2}}-\omega _{c_{1}},\Delta _{23}=\omega _{c_{3}}-\omega
_{c_{2}},$ and $\Delta _{13}=\omega _{c_{3}}-\omega _{c_{1}}$ are the
cavity-frequency detunings.

\begin{figure}[tbp]
\begin{center}
\includegraphics[bb=92 226 737 527, width=15.0 cm, clip]{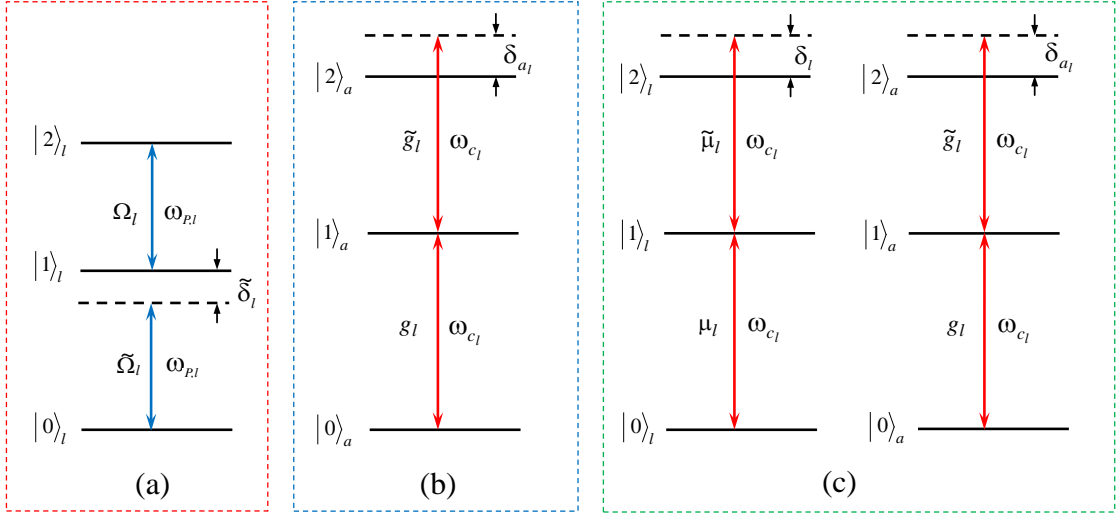} \vspace*{%
-0.08in}
\end{center}
\caption{(Color online) (a) A classical pulse is resonant with the $%
\left\vert 1\right\rangle \leftrightarrow \left\vert 2\right\rangle $
transition of qutrit $l$ with a Rabi frequency $\Omega_{l}$, while
off-resonant with the $\left\vert 0\right\rangle \leftrightarrow \left\vert
1\right\rangle $ transition of qutrit $l$ with a Rabi frequency $\widetilde{%
\Omega }_{l}$ and detuning $\widetilde{\protect\delta }_{l}$ ($l=1,2,3$).
(b) Cavity $l$ is resonant with the $\left\vert 0\right\rangle
\leftrightarrow \left\vert 1\right\rangle $ transition of qutrit $a$ with a
coupling constant $g_l$, while off-resonant with the $\left\vert
1\right\rangle \leftrightarrow \left\vert 2\right\rangle $ transition of
qutrit $a$ with a coupling constant $\widetilde{g_l}$ and detuning $\protect%
\delta _{a_l}$ ($l=1,2,3$). (c) Cavity $l$ is simultaneously resonant with
the $\left\vert 0\right\rangle \leftrightarrow \left\vert 1\right\rangle $
transition of qutrits $l$ and $a$, with a coupling constant $\protect\mu_l$
for qutrit $l$ and a coupling constant $g_l$ for qutrit $a$ ($l=1,2,3$). In
addition, cavity $l$ is off-resonant with the $\left\vert 1\right\rangle
\leftrightarrow \left\vert 2\right\rangle $ transition of qutrits $l$ and $a$%
, with a coupling constant $\widetilde{\protect\mu_l}$ and detuning $\protect%
\delta _{l}$ for qutrit $l$, while a coupling constant $\widetilde{g_l}$ and
detuning $\protect\delta _{a_l}$ for qutrit $a$ ($l=1,2,3$). For the
definition of the detunings $\widetilde{\protect\delta }_{l}$, $\protect%
\delta _{l}$, and $\protect\delta _{a_l}$, refer to the text. }
\label{fig:8}
\end{figure}

(ii) The second basic operation is described by the Hamiltonian $H_{\mathrm{I%
}_{2}}$ in Eq.~(3). In practice, the inter-cavity crosstalk between the
three cavities and the unwanted coupling of cavity $l$ ($l=1,2,3$) with the $%
|1\rangle \leftrightarrow |2\rangle $ transition of qutrit $a$ should be
considered. When these factors are taken into account, the Hamiltonian $H_{%
\mathrm{I}_{2}}$ is modified as

\begin{equation}
\widetilde{H}_{\mathrm{I}_{2}}=\hbar \left( g_{l}a_{l}^{+}\left\vert
0\right\rangle _{a}\left\langle 1\right\vert +\text{h.c.}\right) +\hbar
\left( \tilde{g}_{l}e^{i\delta _{a_{l}}t}a_{l}^{+}\left\vert 1\right\rangle
_{a}\left\langle 2\right\vert +h.c.\right) +\varepsilon ,
\end{equation}%
where the first bracket term represents the resonant interaction of cavity $%
l $ with the $|0\rangle \leftrightarrow |1\rangle $ transition of qutrit $a$%
, while the second bracket term represents the unwanted coupling between
cavity $l$ and the $|1\rangle \leftrightarrow |2\rangle $ transition of
qutrit $a$ with coupling strength $\widetilde{g}_{l}$ and detuning $\delta
_{a_{l}}=\omega _{c_{l}}-\omega _{21,l}^{a}>0$ [Fig.~8(b)]. Here and below,
we define $\omega _{21,l}^{a}$ ($\omega _{10,l}^{a}$) as the $|1\rangle
\leftrightarrow |2\rangle $ ($|0\rangle \leftrightarrow |1\rangle $)
transition frequency of qutrit $a$ in the case when qutrit $a$ interacts
with cavity $l.$ Note that both of $\omega _{21,l}^{a}$ and $\omega
_{10,l}^{a}$ change with the index $l$ (i.e., the qutrit-cavity interaction
switches from one cavity to another). The detuning $\delta _{a_{l}}$ can
also be written as $\delta _{a_{l}}=\omega _{10,l}^{a}-\omega
_{21,l}^{a}=\delta \omega _{l}^{a}$ because of $\omega _{c_{l}}=\omega
_{10,l}^{a}.$ Here, $\delta \omega _{l}^{a}$ is the anharmonicity between
the $|0\rangle \leftrightarrow |1\rangle $ level spacing and the $|1\rangle
\leftrightarrow |2\rangle $ level spacing of qutrit $a$. The last term of
Eq.~(17) represents the inter-resonator crosstalk, which is described by Eq.
(12) given above.

(iii) The third basic operation is described by the Hamiltonian $H_{\mathrm{I%
}_{3}}$ in Eq.~(5). In reality, the inter-cavity crosstalk between the three
cavities and the unwanted coupling of cavity $l$ ($l=1,2,3$) with the $%
|1\rangle \leftrightarrow |2\rangle $ transition of both qutrit $a$ and
qutrit $l$ need to be considered. After taking these factors into account,
the Hamiltonian $H_{\mathrm{I}_{\mathrm{3}}}$ becomes

\begin{eqnarray}
\widetilde{H}_{\mathrm{I}_{3}} &=&\hbar \left( \mu _{l}a_{l}^{+}\left\vert
0\right\rangle _{l}\left\langle 1\right\vert +\text{h.c.}\right) +\hbar
\left( g_{l}a_{l}^{+}\left\vert 0\right\rangle _{a}\left\langle 1\right\vert
+\text{h.c.}\right)  \notag \\
&&+\hbar \left( \widetilde{\mu }_{l}e^{i\delta _{l}t}a_{l}^{+}\left\vert
0\right\rangle _{l}\left\langle 1\right\vert +\text{h.c.}\right)  \notag \\
&&+\hbar \left( \tilde{g}_{l}e^{i\delta _{a_{l}}t}a_{l}^{+}\left\vert
0\right\rangle _{a}\left\langle 1\right\vert +\text{h.c.}\right)  \notag \\
&&+\varepsilon ,
\end{eqnarray}%
where the first (second)\ bracket term in the first line represent the
resonant interaction of cavity $l$ with the $|0\rangle \leftrightarrow
|1\rangle $ transition of qutrit $l$ ($a$), the term in the second line
represent the unwanted coupling between cavity $l$ and the $|1\rangle
\leftrightarrow |2\rangle $ transition of qutrit $l$ with coupling strength $%
\widetilde{g}_{l}$ and detuning $\delta _{_{l}}=\omega _{c_{l}}-\omega
_{21,l}=\omega _{10,l}-\omega _{21,l}=\delta \omega _{l}>0$ [Fig.~8(c)],
while the term in the third line represent the unwanted coupling between
cavity $l$ and the $|1\rangle \leftrightarrow |2\rangle $ transition of
qutrit $a$ with coupling strength $\widetilde{g}_{l}$ and detuning $\delta
_{a_{l}}$ described above. Here, $\omega _{10,l}$ ($\omega _{21,l}$)
represents the $|0\rangle \leftrightarrow |1\rangle $ ($|1\rangle
\leftrightarrow |2\rangle $) transition frequency of qutrit $l.$ We have
applied $\omega _{c_{l}}=\omega _{10,l}$ in writing $\delta _{_{l}}=\delta
\omega _{l}$. Here, $\delta \omega _{l}$ is the anharmonicity between the $%
|0\rangle \leftrightarrow |1\rangle $ level spacing and the $|1\rangle
\leftrightarrow |2\rangle $ level spacing of qutrit $l.$

By considering dissipation and dephasing, the evolution of the system is
determined by the master equation

\begin{eqnarray}
\frac{d\rho }{dt} &=&-i\left[ \widetilde{H}_{I_{k}},\rho \right]
+\sum_{l=1}^{3}\kappa _{a_{l}}\mathcal{L}\left[ a_{l}\right]  \notag \\
&&+\sum_{l=1,2,3,a}\gamma _{12,l}\mathcal{L}\left[ \sigma _{12,l}^{-}\right]
+\gamma _{02,l}\mathcal{L}\left[ \sigma _{02,l}^{-}\right] +\gamma _{01,l}%
\mathcal{L}\left[ \sigma _{01,l}^{-}\right]  \notag \\
&&+\sum_{l=1,2,3,a}\gamma _{2\varphi ,l}\left( \sigma _{22,l}\rho \sigma
_{22,l}-\sigma _{22,l}\rho /2-\rho \sigma _{22,l}/2\right)  \notag \\
&&+\sum_{l=1,2,3,a}\gamma _{1\varphi ,l}\left( \sigma _{11,l}\rho \sigma
_{11,l}-\sigma _{11,l}\rho /2-\rho \sigma _{11,l}/2\right) ,
\end{eqnarray}%
where $\widetilde{H}_{I_{k}}$ (with $k=1,2,3$) are the modified Hamiltonians
$\widetilde{H}_{I_{1}},$ $\widetilde{H}_{I_{2}},$ and $\widetilde{H}_{I_{3}}$%
, $\mathcal{L}\left[ \Lambda \right] =\Lambda \rho \Lambda ^{+}-\Lambda
^{+}\Lambda \rho /2-\rho \Lambda ^{+}\Lambda /2$ (with $\Lambda
=a_{l},\sigma _{12,l}^{-},\sigma _{02,l}^{-},\sigma _{01,l}^{-})$,\ $\sigma
_{12,l}^{-}=\left\vert 1\right\rangle _{l}\left\langle 2\right\vert ,$ $%
\sigma _{02,l}^{-}=\left\vert 0\right\rangle _{l}\left\langle 2\right\vert ,$
$\sigma _{01,l}^{-}=\left\vert 0\right\rangle _{l}\left\langle 1\right\vert
, $ $\sigma _{22,l}=\left\vert 2\right\rangle _{l}\left\langle 2\right\vert $%
, and $\sigma _{11,l}=\left\vert 1\right\rangle _{l}\left\langle
1\right\vert ; $ $\kappa _{a_{l}}$ is the decay rate of cavity $l$ ($l=1,2,3$%
);\ $\gamma _{12,l}$ ($\gamma _{02,l}$) is the energy relaxation rate for
the level $\left\vert 2\right\rangle $ associated with the decay path $%
\left\vert 2\right\rangle \rightarrow \left\vert 1\right\rangle $ ($%
\left\vert 2\right\rangle \rightarrow \left\vert 0\right\rangle $) of qutrit
$l$; $\gamma _{01,l}$ is the energy relaxation rate of the level $\left\vert
1\right\rangle ;$ and $\gamma _{2\varphi ,l}$ ($\gamma _{1\varphi ,l}$) is
the dephasing rate of the level $\left\vert 2\right\rangle $ ($\left\vert
1\right\rangle $) of qutrit $l$ ($l=1,2,3,a$)\textbf{. }

The fidelity of the whole operation is given by $\mathcal{F}=\sqrt{%
\left\langle \psi _{\mathrm{id}}\right\vert \rho \left\vert \psi _{\mathrm{id%
}}\right\rangle },$ where $\left\vert \psi _{\mathrm{id}}\right\rangle $ is
the ideal output state obtained under the theoretical model, while $\rho $
is the final density matrix obtained by numerically solving the master
equation. As an example, we consider an input state of the whole system$%
\frac{1}{2\sqrt{2}}\sum \left\vert i_{1}i_{2}i_{3}\right\rangle \otimes
\left\vert 0\right\rangle _{a}\left\vert 0\right\rangle _{c}$, with $%
i_{1},i_{2},i_{3}\in \left\{ 0,1\right\} .$ Thus, the ideal output state is $%
\left\vert \psi _{\mathrm{id}}\right\rangle =\frac{1}{2\sqrt{2}}\left(
\sum_{i_{1},i_{2},i_{3}\neq 1}\left\vert i_{1}i_{2}i_{3}\right\rangle
-\left\vert 111\right\rangle \right) \otimes \left\vert 0\right\rangle
_{a}\left\vert 0\right\rangle _{c}.$

We now numerically calculate the fidelity. The anharmonicity of the level
spacings for a transmon qutrit can be made to be within $100\sim 600$ MHz
[60]. Note that the anharmonicity of the level spacings of a transmon qutrit
slightly varies during adjusting the qutrit level spacings. Thus, we can
choose $\delta \omega _{l}/2\pi \sim \delta \omega _{a_{l}}/2\pi \sim \delta
\widetilde{\omega }_{l}/2\pi =600$ MHz ($l=1,2,3$). For a TLR, the typical
frequency is $1-10$ GHz. Therefore, we choose $\omega _{c_{1}}/2\pi =5.0$
GHz, $\omega _{c_{2}}/2\pi =6.0$ GHz, and $\omega _{c_{3}}/2\pi =7.0$ GHz,
resulting in $\Delta _{12}/2\pi =\Delta _{23}/2\pi =1$ GHz and $\Delta
_{23}/2\pi =2$ GHz. We choose $\mu _{l}/2\pi =g_{l}/2\pi =g/2\pi =10$ MHz
and $\Omega _{l}/2\pi =\Omega /2\pi =15$ MHz ($l=1,2,3$), which are
available in experiments [61,62]. For a transmon qutrit [63], one has $%
\widetilde{\mu }_{l}\sim \sqrt{2}\mu _{l},$ $\widetilde{g}_{l}\sim \sqrt{2}%
g_{l}$, and $\widetilde{\Omega }_{l}\sim \Omega /\sqrt{2}$. Other parameters
used in the numerical simulations are: (i) $\gamma _{1\varphi ,l}^{-1}=15$ $%
\mu $s$,$ $\gamma _{2\varphi ,l}^{-1}=15$ $\mu $s; (ii) $\gamma
_{01,l}^{-1}=20$ $\mu $s, $\gamma _{12,l}^{-1}=10$ $\mu $s, $\gamma
_{02,l}^{-1}=25$ $\mu $s ($l=1,2,3,a$), and (iii) $\kappa _{a_{l}}^{-1}=10$ $%
\mu $s ($l=1,2,3$). It should be mentioned that the decoherence times chosen
here is a conservative case because the energy relaxation time and the
dephasing time can be made to be on the order of $25-100$ $\mu $s for the
state-of-the-art superconducting transmon devices at the present time
[64-67]. For simplicity, we choose $g_{12}=g_{23}=g_{13}=0.1g$ in our
numerical simulations.

\begin{figure}[tbp]
\begin{center}
\includegraphics[bb=0 0 321 136, width=15.0 cm, clip]{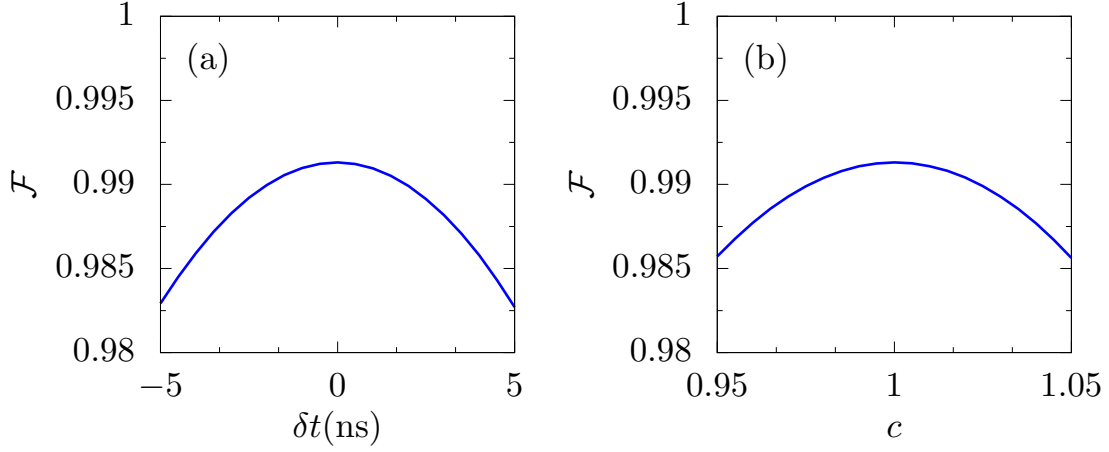} \vspace*{%
-0.08in}
\end{center}
\caption{(Color online) (a) Fidelity versus $\protect\delta t$. Here, $%
\protect\delta t$ is an error, which applies to each typical interaction
time. (b) Fidelity versus $c=\protect\mu_l/g_l$. For the definition of $%
\protect\mu_l$ and $g_l$, see the text.}
\label{fig:9}
\end{figure}

\begin{figure}[tbp]
\begin{center}
\includegraphics[bb=56 0 750 450, width=10.0 cm, clip]{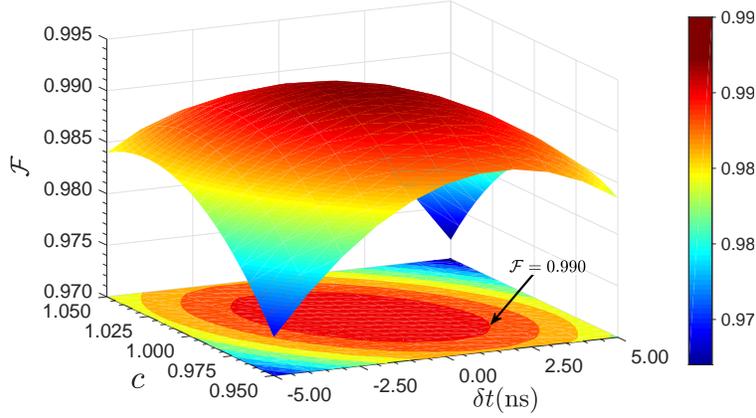} \vspace*{%
-0.08in}
\end{center}
\caption{(Color online) (a) Fidelity versus $\protect\delta t$ and $c$.
Here, $\protect\delta t$ is an error for each typical interaction time. In
addition, $c=\protect\mu_l/g_l$.}
\label{fig:10}
\end{figure}

To investigate the effect of the operation time error on the fidelity, we
apply an error $\delta t$ to each international time as described in the
previous section. With this modification, we numerically calculate the
fidelity and plot Fig.~9(a). Figure~9(a) shows that the fidelity is
sensitive to the operation time error but a high fidelity greater than $%
98.8\% $ can be reached for $\delta t\in \lbrack -5,5]$ ns, i.e.,
corresponding to a $10\%$ error in each typical operation time, which is $%
\sim 50$ ns based on the values of $g$ and $\Omega $ chosen above.

In a realistic situation, it may be a challenge to meet the condition $\mu
_{l}=g_{l}$ ($l=1,2,3$). Thus, we consider a small difference between $\mu
_{l}$ and $g_{l}.$ We set $\mu _{l}=cg_{l}$ with $c\in \left[ 0.95,1.05%
\right] .$ Here, $g_{l}/2\pi =g/2\pi =10$ MHz. We numerically calculate the
fidelity and plot Fig.~9(b). Figure~9(b) demonstrates that for $c\in \left[
0.95,1.05\right] $, the fidelity is greater than $99.1\%$.

We further plot Fig.~10, which shows the fidelity versus the operation time
error $\delta t$ and the coupling-strength ratio $c=\mu _{l}/g_{l}$. From
Fig.~10, one can see that the fidelity can reach 0.99 or greater for $-3$ ns $%
\leq \delta t\leq 3$ ns and $0.97\leq c\leq 1.03$, i.e., the area indicated
by the red circle at the bottom of Fig.~10.

The entire operation time is $\sim 0.5$ $\mu $s for the values of $g$ and $%
\Omega $ given above. Note that the operation time can be shortened by
increasing $g$ and $\Omega .$ For instance, the entire operation time could
be be $\sim 50$ ns if we choose $g/2\pi =100$ MHz and $\Omega /2\pi =150$ MHz%
$.$ However, due to the smaller anharmonicity of the level spacings for a
transmon qutrit, increasing $g$ and $\Omega $ may decrease the fidelity,
because the effect of the unwanted $|0\rangle \leftrightarrow |1\rangle $
coupling (induced by the pulse) and the effect of the unwanted $|1\rangle
\leftrightarrow |2\rangle $ coupling (caused by the cavity mode) become more
apparent as $g$ and $\Omega $ increase.

As discussed in [68,69], as long as the cavities are physically well
separated, the inter-cavity crosstalk coupling strength is $g_{lk}\approx
g_{l}\left( C_{l}/C_{\Sigma }\right) $ ($lk=12,23,13$), where $C_{\Sigma
}=C_{1}+C_{2}+C_{3}+C_{a}$ is the sum of the three coupling capacitances and
the qutrit $a$'s capacitance. For $C_{1},C_{2},C_{3}\sim 1$ fF and $%
C_{\Sigma }\sim 10^{2}$ fF (the typical values in experiments [68,69]), we
have $g_{lk}\leq 0.1g_{l}.$ Note that in the numerical simulations, we set $%
g_{l}=g$. Thus, the condition $g_{lk}=0.1g$ ($lk=12,23,13$) can be easily
satisfied.

For the cavity frequencies given above and $\kappa _{a_{l}}^{-1}=10$ $\mu $%
s, the required quality factors for the three cavities are $Q_{1}=3.1\times
10^{5},$ $Q_{2}=3.8\times 10^{5},$ and $Q_{3}=4.4\times 10^{5}.$ The
required cavity quality factors here are achievable in experiments because
TLRs with a (loaded) quality factor $Q\sim 10^{6}$ have been experimentally
demonstrated [70,71]. Our analysis here implies that high-fidelity
implementation of a three-qubit controlled phase gate, which act on three
qubits distributed in three different cavities, is feasible with present
circuit QED technology.

It is worth noting that the capacitive coupling of a superconducting transmon qutrit
to two one-dimension transmission line resonators has been experimentally implemented [72]. With rapid development
of the circuit QED technology, we think that the setup illustrated in Fig. 7, which consists of three one-dimension transmission line
resonators capacitively coupled to a single transmon qutrit, will soon be realized in experiments.

\begin{center}
\textbf{VI. THREE-QUBIT CONTROLLED PHASE GATE WITH ATOMS USING A SINGLE CAVITY}
\end{center}

The above method can be generalized to implement an $n$-qubit controlled
phase gate with $n+1$ atoms, by using one cavity. To see this, let us consider a
three-qubit case. We will give a detailed discussion on realizing a
three-qubit controlled phase gate described above, using atoms interacting
with a single cavity.

Consider four identical three-level atoms ($1,2,3,a$) and a single cavity.
Each atom has three levels as shown in Fig. 2(b). Suppose that
the transition between the two levels $\left\vert 0\right\rangle $ and $%
\left\vert 1\right\rangle $ of each atom is resonantly coupled to the cavity
with a coupling constant $g$, while the transition between any other two
levels is highly detuned or decoupled from the cavity mode. In addition,
assume that the cavity is initially in the vacuum state. Each atom is
trapped in the periodic potential of a one-dimensional optical lattice and
can be loaded into or moved out of the cavity by translating the optical
lattice [73]. Alternatively, each atom is trapped in an optical dipole trap
and can be loaded into or moved out of the cavity by translating the optical
dipole trap [74].

\begin{figure}[tbp]
\begin{center}
\includegraphics[bb=63 106 576 693, width=15.0 cm, clip]{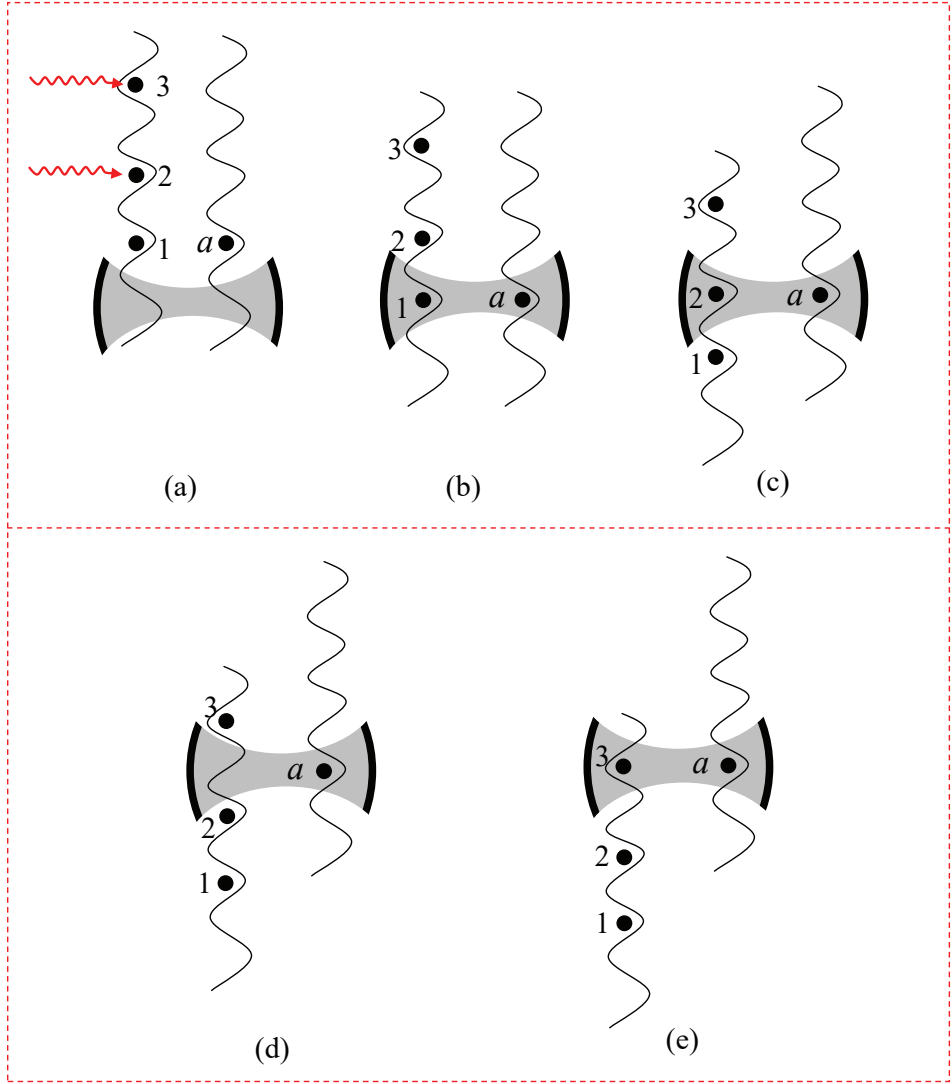} \vspace*{%
-0.08in}
\end{center}
\caption{(Color online) Proposed setup for a three-qubit controlled phase
gate with four identical neutral atoms ($1,2,3,a$) and a cavity. Each atom
can be either loaded into the cavity or moved out of the cavity by
one-dimensional translating optical lattices [73] or an optical dipole trap
[74]. Atoms 1 and 2 play a role of controlled qubits while atom 3 acts as a
target qubit. In addition, atom $a$ is an auxiliary system, which is used
for the coherent manipulation.}
\label{fig:11}
\end{figure}

According to the gate operations described in subsection III A, one can
easily find that the procedure for implementing the three-qubit controlled
phase gate is listed below:

Step 1: Apply a classical pulse to each of atoms ($2,3$), which has an
initial phase $-\pi /2$ and is resonant with the $\left\vert 1\right\rangle
\leftrightarrow \left\vert 2\right\rangle $ transition [Fig. 11(a)]. The
Rabi frequency of the pulse applied to atom $l$ is $\Omega $ ($l=2,3$).
After an interaction time $\tau _{1}=\pi /\left( 2\Omega \right) ,$ the
state $\left\vert 1\right\rangle $ of atom $l$ changes to $\left\vert
2\right\rangle $ according to Eq.~(2).

Step 2: Move atoms $1$ and $a$ to resonance with cavity $1$ for an
interaction time $\tau _{2}=\pi /\left( \sqrt{2}g\right) $ [Fig.~11(b)],
resulting in the state transformation $\left\vert 1\right\rangle
_{1}\left\vert 0\right\rangle _{a}\left\vert 0\right\rangle
_{c_{1}}\rightarrow -\left\vert 0\right\rangle _{1}\left\vert 1\right\rangle
_{a}\left\vert 0\right\rangle _{c_{1}}$ according to Eq.~(6).

Step 3: Move atom 1 out of the cavity but move atom 2 into the cavity and
keep atom $a$ in the cavity to have atoms 2 and $a$ to resonance with cavity
2 for an interaction time $\tau _{3,1}=\pi /\left( \sqrt{2}g\right) $
[Fig.~11(c)], resulting in $\left\vert 0\right\rangle _{2}\left\vert
1\right\rangle _{a}\left\vert 0\right\rangle _{c_{2}}\rightarrow -\left\vert
1\right\rangle _{2}\left\vert 0\right\rangle _{a}\left\vert 0\right\rangle
_{c_{2}}$ according to Eq. (6). Then move atom 2 out of the cavity such that
the state $-\left\vert 1\right\rangle _{2}\left\vert 0\right\rangle
_{a}\left\vert 0\right\rangle _{c_{2}}$ remains unchanged, while let atom $a$
continue resonantly interacting with cavity 2 for an additional interaction
time $\tau _{3,2}=2\pi /g-\pi /\left( \sqrt{2}g\right) $ [Fig.~11(d)]. After
an interaction time $\tau _{3}=\tau _{3,1}+$ $\tau _{3,2}=2\pi /g,$ the
state $\left\vert 1\right\rangle _{a}\left\vert 0\right\rangle _{c_{2}}$
remains unchanged according to Eq.~(4).

Step 4: Move atom $3$ into the cavity and keep atom $a$ in the cavity to
have atoms $3$ and $a$ to resonance with cavity $3$ for an interaction time $%
\tau _{4,1}=\pi /\left( \sqrt{2}g\right) $ [Fig.~11(e)], resulting in $%
\left\vert 0\right\rangle _{3}\left\vert 1\right\rangle _{a}\left\vert
0\right\rangle _{c_{3}}\rightarrow -\left\vert 1\right\rangle _{3}\left\vert
0\right\rangle _{a}\left\vert 0\right\rangle _{c_{3}}$ according to Eq. (6).
Then, move atom $3$ out of the cavity such that the state $-\left\vert
1\right\rangle _{3}\left\vert 0\right\rangle _{a}\left\vert 0\right\rangle
_{c_{3}}$ remains unchanged, while let atom $a$ continue resonantly
interacting with cavity $3$ for an additional interaction time $\tau
_{4,2}=\pi /g-\pi /\left( \sqrt{2}g\right) $ [Fig.~11(d)]. After an
interaction time $\tau _{4}=\tau _{4,1}+$ $\tau _{4,2}=\pi /g$, the state $%
\left\vert 1\right\rangle _{a}\left\vert 0\right\rangle _{c_{3}}$ becomes $%
-\left\vert 1\right\rangle _{a}\left\vert 0\right\rangle _{c_{3}}$ according
to Eq.~(4).

Step 5: Let atom $a$ to resonance with cavity 3 for an interaction time $%
\tau _{5,1}=2\pi /g-\pi /\left( \sqrt{2}g\right) $ [Fig.~11(d)]. Then, keep
atom $a$ in the cavity and move atom $3$ into the cavity to have atoms $a$
and $3$ to resonance with cavity $3$ for an interaction time $\tau
_{5,2}=\pi /\left( \sqrt{2}g\right) $ [Fig.~11(e)]. After the interaction
time $\tau _{5,2},$ we have the state transformation $\left\vert
1\right\rangle _{3}\left\vert 0\right\rangle _{a}\left\vert 0\right\rangle
_{c_{3}}\rightarrow -\left\vert 0\right\rangle _{3}\left\vert 1\right\rangle
_{a}\left\vert 0\right\rangle _{c_{3}}$ according to Eq. (6), while after an
interaction time $\tau _{5}=\tau _{5,1}+\tau _{5,2}=2\pi /g$ the state $%
\left\vert 1\right\rangle _{a}\left\vert 0\right\rangle _{c_{3}}$ remains
unchanged according to Eq. (4).

Step $6$: Move atom 3 out of the cavity but let atom $a$ to resonance with
the cavity for an interaction time $\tau _{6,1}=2\pi /g-\pi /\left( \sqrt{2}%
g\right) $ [Fig.~11(d)]. Then, move atom 2 into the cavity and keep atom $a$
in the cavity to have atoms 2 and $a$ to resonantly interact with the cavity
for an interaction time $\tau _{6,2}=\pi /\left( \sqrt{2}g\right) $
[Fig.~11(c)]. After the interaction time $\tau _{6,2},$ one has the state
transformation $\left\vert 1\right\rangle _{2}\left\vert 0\right\rangle
_{a}\left\vert 0\right\rangle _{c_{2}}\rightarrow -\left\vert 0\right\rangle
_{2}\left\vert 1\right\rangle _{a}\left\vert 0\right\rangle _{c_{2}}$
according to Eq.~(6), while after an interaction time $\tau _{6}=\tau
_{6,1}+\tau _{6,2}=2\pi /g$ the state $\left\vert 1\right\rangle
_{a}\left\vert 0\right\rangle _{c_{2}}$ remains unchanged according to Eq.
(4).

Step $7$: Move atom 2 out of the cavity but move atom 1 into the cavity to
have atoms $1$ and $a$ to resonance with the cavity for an interaction time $%
\tau _{7}=\pi /\left( \sqrt{2}g\right) $ [Fig.~11(b)], resulting in the
state transformation $\left\vert 0\right\rangle _{1}\left\vert
1\right\rangle _{a}\left\vert 0\right\rangle _{c_{1}}\rightarrow -\left\vert
1\right\rangle _{1}\left\vert 0\right\rangle _{a}\left\vert 0\right\rangle
_{c_{1}}$ according to Eq.~(6).

Step 8$:$ Move atoms 1 and $a$ out of the cavity [Fig. 11(a)]. Then, apply a
classical pulse to each of atoms ($2,3$), which has an initial phase $\pi /2$
and is resonant with the $\left\vert 1\right\rangle \leftrightarrow
\left\vert 2\right\rangle $ transition of the atoms. The Rabi frequency of
the pulse applied to atom $l$ is $\Omega $ ($l=2,3$). After a pulse duration
$\tau _{8}=\pi /\left( 2\Omega \right) ,$ the state $\left\vert
2\right\rangle $ of atom $l$ changes to $\left\vert 1\right\rangle $
according to Eq.~(2).

The total operation time is $\tau =\pi /\Omega +\left( \sqrt{2}+7\right) \pi
/g+10\tau _{\mathrm{m}}.$ Here, $\tau _{\mathrm{m}}$ is the typical time
required for moving atoms into or out of the cavity. For the proposal to
work, the $\tau $ should be much smaller than the energy relaxation time of
level $\left\vert 1\right\rangle $ or $\left\vert 2\right\rangle $, such
that the decoherence induced due to the spontaneous decay of the level $%
\left\vert 1\right\rangle $ or $\left\vert 2\right\rangle $ is negligible.
In addition, it is noted that during the gate operation, the longest single
atom-cavity interaction time is $2\pi /g,$ which should be much shorter than
the cavity decay time $\kappa ^{-1}$, so that the cavity dissipation is
negligible. In principle, these conditions can be satisfied by choosing a
cavity with a high quality factor $Q$ and atoms with a sufficiently long
energy relaxation time.

To investigate the experimental feasibility of this proposal, let us
consider Rydberg atoms with principal quantum numbers 49--51 (respectively
corresponding to the levels $\left\vert 0\right\rangle ,$ $\left\vert
1\right\rangle ,$ and $\left\vert 2\right\rangle $). The $\left\vert
0\right\rangle \leftrightarrow \left\vert 1\right\rangle $ transition
frequency is $\omega _{0}/2\pi \sim 51.1$ GHz, the energy relaxation time $%
T_{\mathrm{rad}}$ of the level $\left\vert 1\right\rangle $ or $\left\vert
2\right\rangle $ is on the order of $10^{-2}$ s [75], and the coupling
constant is $g=2\pi \times 50$ KHz [75,76]. In addition, choose $\Omega
=2\pi \times 50$ KHz. With the choice of these parameters, the time needed
for the entire operation is $\tau \sim $ 104 $\mu $s for $\tau _{%
\mathrm{m}}\sim 1$ $\mu $s [4], which is much shorter than $T_{\mathrm{rad}}$
by two orders of magnitude. The cavity mode frequency is $\omega _{c}/2\pi
\sim 51.1$ GHz, thus the lifetime of the cavity photon is $%
T_{c}=Q/\omega _{c}\sim 200$ $\mu $s for a cavity with $Q=6.4\times 10^{7}$,
which is much larger than the longest single atom-cavity interaction time $2\pi
/g\sim 20$ $\mu $s$.$ Note that optical cavities with a high $Q\sim 10^{10}$ or
greater have been demonstrated in experiments [77,78]. Thus, the present
proposal might be realizable using current cavity QED setups.

\begin{center}
\textbf{VII. CONCLUSION}
\end{center}

We have presented an efficient method to realize an n-qubit controlled phase
gate or an n-qubit Toffoli gate with n qubits distributed in different
cavities. As shown above, the method has the following advantages: (i) Only
one auxiliary qutrit is used to couple the cavities and no other auxiliary
system is required, thus the hardware resources is significantly reduced;
(ii) The gate implementation is deterministic since no measurement is
needed; (iii) The gate can in principle be performed fast for a small number
of qubits, because of only employing resonant interaction; and (iv) By using
the conventional gate-decomposing protocols to construct the
multi-control-qubit gate, the number of required basic gates (single-qubit
or two-qubit gates) drastically increases as the number of qubits increases.
In a stark contrast, as shown above, the number of basic operations,
required by our proposal, only increases \textit{linearly} with the number
of qubits. Thus, the gate realization by this proposal is greatly simplified
when compared to using conventional gate decomposing protocols.

We have further investigated the experimental feasibility of implementing
the proposed gate for a three-qubit case, based on circuit QED. Numerical
simulations show that high-fidelity implementation of a three-qubit
controlled phase gate, which is executed on three qubits distributed in
three different cavities, is feasible within current circuit QED technology.
The method is quite general and can be applied to realize the proposed gate
with qubits distributed in different cavites, in various physical systems,
such as natural atoms or artificial atoms (e.g., quantum dots, NV centers,
various superconducting qutrits, etc.) distributed in different cavities.

Finally, it is noted that the method can be applied to implement a
multiqubit controlled phase gate with atoms and a single cavity. As an
example, we have explicitly shown how to realize a three-qubit
controlled phase gate by loading atoms into or moving atoms out of the
cavity, and have analyzed the experimental feasibility.

Before ending conclusion, we should mention that the architecture illustrated in Fig. 1 or Fig. 2(a) would offer more flexibility
for a circuit-QED system, when compared with a cavity-QED system consisting of atoms and optical cavities. For atoms, adjusting the atomic level spacings is usually difficult and slow in practice, and loading an atom into more than one cavity simultaneously is challenging
and has not been reported in experiments. Also, the speed of adjusting the frequency of an optical cavity is quite slow in experiments. In contrast, as discussed previously, the level spacings of superconducting qutrits can be rapidly adjusted within a few nanoseconds, and the frequency of a microwave cavity or resonator can be quickly tuned in 1--3 ns. In addition,
the ancillary superconducting qutrit can be used to couple cavities or resonators via capacitors. By combining these degrees of freedom,
several types of important couplings can be achieved, such as: (i) the coupling of the ancillary qutrit with the selective cavities;
(ii) the coupling between the selective cavities; and (iii)
the coupling among qubits distributed in different cavities. These couplings
are necessary in quantum gate implementation, entanglement generation, and quantum state transfer
with photonic qubits or superconducting qubits distributed in different cavities, which are of importance in
large-scale quantum information processing based on a multi-cavity circuit QED system.

\begin{center}
\textbf{ACKNOWLEDGMENTS}
\end{center}

This work was supported in part by the NKRDP of China (Grant No.
2016YFA0301802), the National Natural Science Foundation of China under
Grant Nos. [11074062, 11374083, 11774076], and the Zhejiang Natural Science
Foundation under Grant No. LZ13A040002. This work was also supported by the
funds from Hangzhou City for the Hangzhou-City Quantum Information and
Quantum Optics Innovation Research Team.

\bigskip

\end{document}